\documentclass[11pt, scriptaddress, article,showpacs,showkeys,nofootinbib]{revtex4-1}

	\usepackage{amsmath}
           \usepackage{mathrsfs}
	\usepackage{amssymb}
\usepackage{tikz}
\usetikzlibrary{arrows}
	\usepackage{makeidx}
	\usepackage{amsfonts}
	\usepackage[ansinew]{inputenc}
	\usepackage[usenames, dvipsnames]{pstricks}
	\usepackage{pst-grad} % For gradients
	\usepackage{pst-plot} % For axes
	\usepackage[colorlinks, hyperindex]{hyperref}
	\usepackage{colortbl}
	\hypersetup
	{
		colorlinks, %
		citecolor=blue, %
		linkcolor=red, %
		urlcolor=blue, %
	}
\usepackage{enumerate}

\allowdisplaybreaks[1]
\definecolor{palegreen}{rgb}{0.59,0.985,0.596}
\definecolor{lightyellow}{rgb}{1,1,0.4}
\definecolor{lightred}{rgb}{0.95,0.5,0.50}
\definecolor{lavender}{rgb}{0.9,0.9,0.98}

\makeindex

\begin{document}

\title{Weyl Consistency Conditions in Non-Relativistic
    Quantum Field Theory}

\author{Sridip Pal} 
\email{srpal@ucsd.edu; sridippaliiser@gmail.com}
\author{Benjam\'\i{}n Grinstein}
\email{bgrinstein@ucsd.edu}
\affiliation{Department of Physics\\
University of California, San Diego\\
9500 Gilman Drive, La Jolla, CA 92093, USA}

\begin{abstract}
  Weyl consistency conditions have been used in unitary relativistic
  quantum field theory to impose constraints on the renormalization
  group flow of certain quantities. We classify the Weyl anomalies
  and their renormalization scheme ambiguities for generic
  non-relativistic theories in $2+1$ dimensions with anisotropic
  scaling exponent $z=2$; the extension to other values of $z$ are
  discussed as well. We give the consistency conditions among these
  anomalies. As an application we find several candidates for a
  $C$-theorem. We comment on possible candidates for a $C$-theorem in
  higher dimensions.
\end{abstract}
%\pacs{To be Written}
\maketitle
\section{Introduction}
Aspects of the behavior of systems at criticality are accessible
through renormalization group (RG) methods. Famously, most critical
exponents are determined by a few anomalous dimensions of
operators. However, additional information, such as dynamical (or
anisotropic) exponents and amplitude relations can be accessed via
renormalization group methods near but not strictly at
criticality. Far away from
critical points there are often other methods, {\it e.g.}, mean field
approximation, that can give more detailed information. The
renormalization group used away from critical points can valuably
bridge the gap between these regions.

Systems of non-relativistic particles at unitarity, in which the
$S$-wave scattering length diverges, $|a|\rightarrow \infty$, exhibit
non-relativistic conformal symmetry. Ultracold atom gas experiments
have renewed interest in study of such theories. In these experiments
one can freely tune the $S$-wave scattering length along an RG flow
\cite{R,Z}: at $a^{-1}=-\infty$ the system is a BCS superfluid while
at $a^{-1}=\infty$ it is a BEC superfluid. The BCS-BEC crossover, at
$a^{-1}=0$, is precisely the unitarity limit, exhibiting conformal
symmetry. This is a regime where universality is expected, with
features independent of any microscopic details of the atomic
interactions \cite{misc}. Other examples of non-relativistic systems
with accidentally large scattering cross section include few nucleon
systems like the deuteron \cite{kaplan} and several atomic systems,
including 
${ }^{85}Rb$\cite{roberts},${}^{138}Cs$ \cite{chin},
${}^{39}K$ \cite{loftus}. 

In the context of critical dynamics the response function exhibits
dynamical scaling. This is characterized by a dynamical scaling
exponent which characterizes anisotropic scaling in the time
domain. There has been recent interest in anisotropic scaling in
systems that are non-covariant extensions of relativistic systems.   The ultraviolet divergences in quantized Einstein gravity are softened if the theory is modified by
inclusion of higher derivative terms in the Lagrangian. Since time derivatives higher than order 2
lead to the presence of ghosts,\footnote{Generically, the $S$-matrix in models with ghosts is not
  unitary. However, under certain conditions on the spectrum of ghosts and the nature of their
  interactions, a unitary $S$-matrix is
  possible~\cite{Lee:1969fy,Lee:1970iw,Grinstein:2007mp,Grinstein:2008bg}. In theories of gravity
  Hawking and Hertog have proposed that ghosts lead to unitarity violation at short distances, and 
  unitarity is a long-distance emergent phenomenon~\cite{Hawking:2001yt}.}  Horava suggested
extending Einstein gravity by terms with higher spatial derivatives but only order-2 time
derivatives~\cite{horava1}. The mismatch in the number of spatial versus time derivatives is a
version of anisotropic scaling, similar to that found in the non-relativistic context.  This has
motivated studies of extensions of relativistic quantum field theories that exhibit anisotropic
scaling at short distances. Independently, motivated by the study of Lorentz violating theories of
elementary particle interactions~\cite{Anselmi:2007ri}, Anselmi found a critical point with exact
anisotropic scaling, a so-called Lifshitz fixed point, in his studies of renormalization properties
of interacting scalar field theories~\cite{Anselmi:2008ry}; see
Refs.~\cite{Anselmi:2008bq,Anselmi:2008bs} for the case of gauge
theories. Anomalous breaking of anisotropic scaling symmetry in the quantum Lifshitz model
has been studied in Ref.~\cite{Baggio:2011ha, arav1, arav2, auzzi, kristan}; see also
Ref.~\cite{Griffin:2011xs} for an analysis using holographic methods. 

Wess-Zumino consistency conditions for Weyl transformations have been used in
  unitary relativistic quantum field theory to impose constraints on the renormalization
  group flow of Weyl anomalies~\cite{Osborn:1991gm}.  In 1+1 dimensions a combination of
  these anomalies gives Zamolodchikov's $C$-function \cite{Zamo}, that famously decreases
monotonically along flows towards long distances, is stationary at fixed points and equals
the central charge of the 2D conformal field theory at the fixed point boundaries of the
flow. Weyl consistency conditions can in fact be used to recover this result
\cite{Osborn:1991gm}. Along the same lines, in 3+1 dimensions Weyl consistency conditions
can be used to show that a quantity $\tilde a$ satisfies
\begin{equation}
\label{eq:a-thm}
\mu\frac{ d\tilde{a}}{d\mu}=\mathcal{H}_{\alpha\beta}\beta^{\alpha}\beta^{\beta}
\end{equation}
where $\mu$ is the renormalization group scale, increasing towards
short distances. The equation shows that at fixed points, characterized
by $\mu\, dg^{\alpha}/d\mu \equiv \beta^\alpha=0$, $\tilde a$ is
stationary. It can be shown in perturbation theory that
$\mathcal{H}_{\alpha\beta} $ is a positive definite symmetric
matrix \cite{Osborn2}. By construction the quantity $\tilde a$ is, at fixed points, the
conformal anomaly $a$ of Cardy, associated with the Euler density
conformal anomaly when
the theory is placed in a curved background~\cite{Cardy:1988cwa}. This is then a
4-dimensional generalization of Zamolodchikov's $C$ function, at least
in perturbation theory. Going beyond 4 dimensions, Weyl consistency
conditions can be used to show that in $d=2n$ dimensions there is a
natural quantity that satisfies~\eqref{eq:a-thm}, and that this
quantity is at fixed points the anomaly associated with the
$d$-dimensional Euler density~\cite{Grinstein:2013cka}.
Concerns about the viability of a $C$-theorem in 6-dimensions were
raised by explicit computations of 
``metric'' $\mathcal{H}_{\alpha\beta} $ in perturbation
theory~\cite{Grinstein:2014xba,Grinstein:2015ina,Osborn:2015rna}. However
it was discovered in Ref.~\cite{Stergiou:2016uqq} that there exists a one
parameter family of extensions of the the quantity $\tilde a$ of
Ref.~\cite{Grinstein:2013cka} that obey  a  $C$-theorem
perturbatively. 

Weyl consistency conditions can also be used to constrain anomalies
in non-relativistic field theories. The constraints imposed at fixed
points have been studied in Ref.~\cite{Baggio:2011ha} for models with anisotropic scaling
exponent $z=2$ in 2-spatial
dimensions; see Refs.~\cite{Adam:2009gq,Gomes:2011di} for studies of the Weyl
  anomaly at  $d=4, z=3$ and $d=6$. Here we investigate
constraints imposed along renormalization group flows. We recover the
results of~\cite{Baggio:2011ha}  by approaching the critical points along the
flows. As mentioned above, there are questions that can only be
accessed through the renormalization group methods applied to flows,
away from fixed points. The additional information obtained from
consideration of Weyl consistency conditions on flows  can be used to ask
a number of questions. For example, we may ask if there is a suitable
candidate for a $C$-theorem. 

A related issue is the possibility of recursive renormalization group
flows. Recursive flows in the perturbative regime have been found in
several examples in $4-\epsilon$ and in 4 dimensional relativistic
quantum field theory
\cite{Fortin:2011ks,Fortin:2011sz,Fortin:2012ic,Fortin:2012cq,Fortin:2012hc,Fortin:2012hn}. Since
Weyl consistency conditions imply $\tilde a$ does not increase along
RG-flows it must be that $\tilde a$ remains constant along recursive
flows. This can be shown directly, that is, without reference to the
monotonicity of the flow; see~\cite{Fortin:2012hn}. In fact one can
show that on recursive flows all physical quantities, not just
$\tilde a$, remain constant: the recursive flow behaves exactly the
same as a single fixed point.  This is as it should be:
    the monotonicity of the flow of $a$ implies that limit cycles do
    not exist in any physically meaningful sense
    \cite{Komargodski:2011vj,Luty:2012ww}; in fact, they may be
    removed by a field and coupling constant redefinition.  However,
it is well known that bona-fide renormalization group limit cycles
exist in some non-relativistic
theories~\cite{efimov,PhysRev.47.903,PhysRevLett.89.230401}. The
$C$-theorem runs afoul of limit-cycles, and an immediate question then
is what invalidates it in models that exhibit recursive flows?  Our
analysis indicates some potential candidates for $C$-theorems but does
not show whether generically the ``metric''
$\mathcal{H}_{\alpha\beta} $ has definite sign. The question of under
what conditions the metric has definite sign, precluding recursive
flows, is left open for further investigation.

The paper is organized as follows. In Sec.\ref{sec:gen} we set-up the computation, using a
background metric and space and time dependent coupling constants that act as sources of
marginal operators. In the section we also clarify the relation between the dynamical
exponent and the classical anisotropic exponent. We then use this formalism in
Sec.~\ref{sec:2dimz2} where we analyze the consistency conditions for the case of
2-spatial dimensions and anisotropic exponent $z=2$. The Weyl consistency conditions and
scheme dependent ambiguities are lengthy, so they are collected in
Apps.~\ref{app:conditions22} and \ref{app:anom-ambig}. In Sec.~\ref{sec:allz} we explore
the case of arbitrary $z$, extending some of the results of the previous section and in
Sec.~\ref{sec:alld} we propose a candidate $C$-theorem for any even spatial dimension. We
offer some general conclusions and review our results in Sec.~\ref{sec:conc}. There is no
trace anomaly equation for the case of zero spatial derivatives, that is, particle quantum
mechanics; we comment on this, and present a simple but useful  theorem that does apply in this
case, in the final appendix, App.~\ref{app:1D}.

\section{Generalities}
\label{sec:gen}
We consider non-relativistic (NR) field theories with point-like
interactions. Although not necessary for the computation of Weyl consistency conditions,
it is convenient to keep in mind a Lagrangian description of the model. The Lagrangian
density $\mathcal{L}=\mathcal{L}(\phi,m,g)$ is a function of fields $\phi(t,\vec x)$, mass
parameters $m$ and coupling constants $g$ that parametrize interaction strengths. We restrict our
attention to models for which the action integral,
\[
S[\phi(\vec x,t)]=\int dt\,d^d\!x\, \mathcal{L}
\]
remains invariant under the rescaling
\begin{equation}
\label{Ascaling}
\vec x \mapsto \lambda \vec x,\qquad t \mapsto \lambda^{z} t,
\end{equation}
that is, 
\[
S[\lambda^{\Delta}\phi(\lambda\vec x,\lambda^z t)]=S[\phi(\vec x,t)]\,.
\]
Here $\Delta$ is the matrix of canonical dimensions of the fields
$\phi$. In a multi-field model the anisotropic scaling exponent $z$ is
common to all fields. Moreover, assuming that the kinetic term in
$\mathcal L$ is local, so that it entails powers of derivative
operators, $z$ counts the mismatch in the number of time derivatives
and spatial derivatives. In the most common cases there is a single
time derivative and $z$ spatial derivatives so that $z$ is an
integer. 

For a simple example, useful to keep in mind for orientation,  the action for a
single complex scalar field with anisotropic scaling $z$ in $d$ dimensions is
given by  
\begin{equation}
\label{eq:simpleEx}
S=\int dt\,d^dx\;\left[im\,\phi^*\overleftrightarrow{\partial_t}\phi-
  \vec\nabla_{i_1}\!\!\cdots\vec\nabla_{i_{z/2}}\phi^* \vec\nabla_{i_1}\!\!\cdots\vec\nabla_{i_{z/2}}\phi-gm^{z/d}|\phi|^{2N}\right]\,,
\end{equation}
where $z$ is an even integer so that the Lagrangian density is
local. If $N=1+z/d $ the scaling property \eqref{Ascaling} holds with
$\Delta = d/2$ (alternatively, if $N\in \mathbb{Z}$, then
$z=d(N-1)\in d\mathbb{Z}$). When \eqref{Ascaling} holds the coupling
constant $g$ is dimensionless. The mass parameters $m$ have dimensions of $T/L^z$, where $T$ and $L$ are
time and space dimensions, respectively.  One may use the mass parameter to
measure time in units of $z$-powers of length, and this can be
implemented by absorbing $m$ into a redefinition, $t=m\hat t$. In
multi-field models one can arbitrarily choose one of the masses to
give the conversion factor and then the independent mass ratios are
dimensionless parameters of the model. In models that satisfy the
scaling property~\eqref{Ascaling}, these mass ratios together with the
coefficients of interaction terms comprise the set of dimensionless
couplings that we denote by $g^\alpha$ below.

The above setup is appropriate for studies of, say, quantum criticality. However the
calculations we present are applicable to studies of thermal systems in equilibrium since
the imaginary time version of the action integral is equivalent to an energy functional in
$d+1$ spatial dimensions. Taking $t=-iy$ in the example of Eq.~\eqref{eq:simpleEx} the
corresponding energy integral is 
\[
H=\int dy\,d^dx\;\left[ m\,\phi^*\overleftrightarrow{\partial_y}\phi+
  \vec\nabla_{i_1}\!\!\cdots\vec\nabla_{i_{z/2}}\phi^*
  \vec\nabla_{i_1}\!\!\cdots\vec\nabla_{i_{z/2}}\phi 
  +gm^{z/d}|\phi|^{2N}\right]\,.
\]

The short distance divergences encountered in these models need to be
regularized and renormalized. Although our results do not depend
explicitly on the regulator used, it is useful to keep in mind a
method like dimensional regularization that retains most symmetries
explicitly. Thus we consider NR field theories in $1+n$ dimensions,
where the spatial dimension $n=d-\epsilon$, with $d$ an
integer. Dimensional regularization requires the introduction of a
parameter $\mu$ with dimensions of inverse length, $L^{-1}$. Invariance
under \eqref{Ascaling} is then broken, but can be formally recovered
by also scaling $\mu$ appropriately, $\mu\mapsto
\lambda^{-1}\mu$. For an example, consider the dimensionally regularized
version of~\eqref{eq:simpleEx}:
\begin{equation}
\label{eq:simpleExDimReg}
S[\phi_0(\vec x,t);\mu]=\int dt\,d^nx\;\left[im_0\,\phi_0^*\overleftrightarrow{\partial_t}\phi_0-
  \vec\nabla_{i_1}\!\!\cdots\vec\nabla_{i_{z/2}}\phi_0^* \vec\nabla_{i_1}\!\!\cdots\vec\nabla_{i_{z/2}}\phi_0-gZ_gm_0^{z/d}\mu^{k\epsilon}|\phi_0|^{2N}\right]\,.
\end{equation}
We have  written this in terms of bare field and mass, $\phi_0$ and
$m_0$, and have given the bare coupling constant explicitly in terms of the
renormalized one, $g_0=\mu^{k\epsilon}Z_gg$. The coefficient
$k=N-1=z/d$ is dictated by dimensional analysis. It follows that 
\begin{equation}
\label{eq:SAnomScaling}
S[\lambda^{n/2}\phi_0(\lambda\vec x,\lambda^z t);\lambda^{-1}\mu]=S[\phi_0(\vec x,t);\mu]
\end{equation}

In order to study the response of the system to sources that couple to
the operators in the interaction terms of the Lagrangian, we
generalize the coupling constants $g^\alpha$ to functions of space and time
$g^\alpha(t,\vec x)$. One can then obtain correlation functions of these
operators by taking functional derivatives of the partition function
with respect to the space-time dependent couplings, and then setting
the coupling functions to constant values. Additional operators of
interest are obtained by placing these systems on a curved background,
with metric $\gamma_{\mu\nu}(t,\vec x)$. One can then obtain
correlations including components of the stress-energy tensor by
taking functional derivatives with respect to the metric and
evaluating these on a trivial, constant metric. For example, we then
can define the components of the symmetric quantum stress energy
tensor and finite composite operators in the following way:
\begin{equation}
\label{eq:TOgiven}
T_{\mu\nu}=\frac2{\sqrt{\gamma}}\frac{\delta S_0}{\delta \gamma^{\mu\nu}}\,\qquad \left[\mathcal{O}_{\alpha}\right]=\frac1{\sqrt{\gamma}}\frac{\delta S_0}{\delta g^{\alpha}}
\end{equation}
The square bracket notation in the last term indicates that these are
finite operators, possibly differing from
$\mathcal{O}_\alpha=\partial\mathcal{L}/\partial g^\alpha$ by a
total derivative term. 

Time plays a special role in theories with anisotropic scaling
symmetry. Hence, it is useful to assume the background space-time, in
addition to being a differential manifold $\mathcal{M}$, carries an
extra structure --- we can foliate the space-time with a foliation of
co-dimension $1$. This can be thought of a topological structure on
$\mathcal{M}$ \cite{horava1}, before any notion of Riemannian metric
is introduced on such manifold. Now the co-ordinate transformations
that preserve the foliation are of the form:
\begin{equation}
t\mapsto\tau(t),\ x^{i}\mapsto \xi^{i}(\vec{x},t)
\end{equation}
We will also assume the space-time foliation is topologically given by
$\mathcal{M}=R \times \Sigma$.  The foliation can be given Riemannian
structure with three basic objects: $h_{ij}$, $N_{i}$ and $N$. This is
the ADM decomposition of the metric --- one can generally think as writing
the metric in terms of lapse and shift functions, $N(t,\vec x)$ and
$N_i(t,\vec x)$, and a metric on spatial sections, $h_{ij}(t,\vec x)$:
\begin{equation}
ds^{2}=\gamma_{\mu\nu}dx^{\mu}dx^{\nu}=N^{2}dt^{2}+2N_idtdx^i-h_{ij}dx^{i}dx^{j}
\end{equation}
Here and below the latin indices run over spatial coordinates,
$i,j=1,\ldots,d$.  We assume invariance of the theory under 
foliation preserving diffeomorphisms. 
In a non-relativistic set up, it is convenient to remove the shift $N^{i}$ by a foliation preserving map
$t\mapsto \tau(t)$ and $x^{i}\mapsto \xi^{i} (\vec{x},t)$.  The metric
is then given by
\begin{equation}
ds^{2}=\gamma_{\mu\nu}dx^{\mu}dx^{\nu}=N^{2}dt^{2}-h_{ij}dx^{i}dx^{j}
\end{equation}
Once the shift functions are  removed the restricted set of
diffeomorphisms that do not mix space and time are allowed,  $t\rightarrow \tau(t)$ and
$x^{i}\rightarrow \xi^{i}(x)$, so that
$N^{i}=0$ is preserved.

In Euclidean space, the generating functional of connected Green's functions $W$ is given by
\begin{equation}
\label{eq:W}
e^W=\int [d\phi]\ e^{-S_0-\Delta S}\,.
\end{equation}
The action integral for these models is generically of the form
\begin{equation}
S_0=\int dt\,d^n\!x\,N\sqrt{h} \mathcal{L}_0\,,
\end{equation}
where $h=\det(h_{ij})$. We have denoted by $\mathcal{L}_0$ the
Lagrangian density with bare fields and couplings as arguments; these
are to be expressed in terms of the renormalized fields and couplings,
so as to render the functional integral finite. 
The  term $\Delta S$ contains additional counter-terms that
are solely functionals of $g^\alpha$ and $\gamma_{\mu\nu}$ that are
also required in order to render $W$ finite. 
In a curved background the scaling 
\eqref{Ascaling} can be rephrased in terms of a transformation of the
metric,
\begin{equation}
\label{eq:metricScalingConst}
N(\vec x, t)\mapsto \lambda^z N(\vec x, t)\,,\qquad h_{ij}(\vec x, t)\mapsto \lambda^2 h_{ij}(\vec x, t)\,.
\end{equation}
Then the generalization of the formal invariance of
Eq.~\eqref{eq:SAnomScaling} is 
\begin{equation}
\label{eq:SAnomScalingCurved} 
S_0[\lambda^z N(\vec x, t),\lambda^2 h_{ij}(\vec x,
t),\lambda^{\Delta_0} \phi_0(\vec x, t);\lambda^{-1}\mu]=
S_0[N(\vec x, t),h_{ij}(\vec x, t), \phi_0(\vec x, t);\mu]
\end{equation}
for a suitable matrix of canonical dimensions $\Delta_0$ of the bare
fields (appropriate to $n=d-\epsilon$ spatial dimensions).

We assume that when introducing a curved background the action integral is suitably
modified so that the formal symmetry of Eq.~\eqref{eq:SAnomScalingCurved} holds locally,
that is, it holds when replacing $\lambda\to \exp(-\sigma(\vec x, t))$. The modification
to the action integral consists of additional terms that couple the fields $\phi$ to the
background curvature.  

For example, the model in Eq.~\eqref{eq:simpleExDimReg} for
$z=2$ is modified to include, in addition to coupling to a background metric, additional
terms
\begin{equation*}
\int dt\,d^nx\,N\sqrt{h}\,\left[im_0\xi_K \phi_0^*\phi_0K+\xi_{N\phi}\left(\phi_{0}^{*}\frac{\partial_{i}N}{N}\partial^{i}\phi_{0}+\phi_{0}\frac{\partial_{i}N}{N}\partial^{i}\phi_{0}^{*}\right)+ \xi_{NN}\frac{\partial_{i}N}{N}\frac{\partial^{i}N}{N}\phi_{0}^{*}\phi_{0} +\xi_{R}R\phi_{0}^{*}\phi_{0}\right]\,.
\end{equation*}
Here $K_{ij}=\frac12\partial_th_{ij}/N$ is the extrinsic curvature of the $t=$~constant
hypersurfaces in the $N^i=0$ gauge and $K=h^{ij}K_{ij}$ (with $h^{ij}$ the inverse of the
metric $h_{ij}$),  and $R$ is the $d$-dimensional Ricci scalar for the metric $h_{ij}$.
Under the transformation \eqref{eq:metricScalingConst} with $\lambda=\exp(-\sigma)$ one
has $K\to e^{2\sigma}(K+n \partial_t\sigma/N)$, $R\to e^{2\sigma}(R+2(n-1)\nabla^{2}\sigma -(n-1)(n-2)\nabla_{i}\sigma\nabla^{i}\sigma)$ and
$N\to e^{-2\sigma}N$, so that choosing $\xi_K=1/2$ and ensuring
 \begin{align}
\label{con}
 2(n-1)\xi_{R}+2\xi_{N\phi}+\frac{n}{2}=0 \qquad (n+2)\xi_{N\phi}-4\xi_{NN}+\frac{n}{2}=0,
 \end{align}
 the action integral remains invariant. Thus, we have a one parameter family of parameters
 that preserves invariance of the action under anisotropic scaling. For arbitrary even $z$
 and arbitrary spatial dimension $n$,  in the
 example~\eqref{eq:simpleExDimReg} we
 first integrate by parts the spatial covariant
 derivatives:
\[
 \vec\nabla_{i_1}\!\!\cdots\vec\nabla_{i_{z/2}}\phi^*
 \vec\nabla_{i_1}\!\!\cdots\vec\nabla_{i_{z/2}}\phi
\to (-1)^{z/2}\phi^*(\nabla^2)^{z/2}\phi\,.
\]
Then   we replace the operator $(\nabla^2)^{\frac{z}{2}}$ by  $\mathcal{O}^{(n+2z-4)}\mathcal{O}^{(n+2z-8)}\cdots\mathcal{O}^{(n+4)}\mathcal{O}^{(n)}$ with $\mathcal{O}^{(p)}$  defined as
\begin{align}
\mathcal{O}^{(p)}\equiv \left[\nabla^{2}-\frac{p}{4(n-1)}R+ \frac{2+p-n}{z}\frac{\partial_{i}N}{N}h^{ij}\partial_{j}+\frac{n}{4z^{2}}(2+p-n)\frac{\partial_{i}N}{N}h^{ij}\frac{\partial_{j}N}{N}\right]
\end{align}
Under $h_{ij}\to e^{-2\sigma}h_{ij}$, $N\to e^{-z\sigma}N$ and $\psi \to
e^{\frac{p}{2}\sigma}\psi$, this operator transform covariantly, in the sense that 
\begin{align}
\mathcal{O}^{(p)}\psi \to e^{(\frac{p}{2}+2)\sigma}\mathcal{O}^{(p)}\psi\,.
\end{align}
Hence, under the Weyl rescaling $h_{ij}\to e^{-2\sigma}h_{ij}$, $N\to e^{-z\sigma}N$ and $\phi \to
e^{\frac{n}{2}\sigma}\phi$ we have following, transforming covariantly
\begin{align}
\phi_{0}^{*}\mathcal{O}^{(n+2z-4)}\mathcal{O}^{(n+2z-8)}\cdots\mathcal{O}^{(n+4)}\mathcal{O}^{(n)}\phi_{0} \to e^{(n+z)\sigma} \phi_{0}^{*}\mathcal{O}^{(n+2z-4)}\mathcal{O}^{(n+2z-8)}\cdots\mathcal{O}^{(n+4)}\mathcal{O}^{(n)}\phi_{0}
\end{align}
For $z=2$, this construction gives 
\begin{align}
N\sqrt{h} \phi_{0}^{*}\mathcal{O}^{(n)}\phi_{0}=N\sqrt{h} \phi_{0}^{*}\left[\nabla^{2}-\frac{n}{4(n-1)}R+\frac{\partial_{i}N}{N}h^{ij}\partial_{j}+\frac{n}{8}\frac{\partial_{i}N}{N}h^{ij}\frac{\partial_{j}N}{N}\right]\phi_0\\
 = N\sqrt{h}\left[-\partial_{i}\phi_{0}^{*}\partial^{i}\phi_{0}-\frac{n}{4(n-1)}R\phi_{0}^{*}\phi_{0}+\frac{n}{8}\frac{\partial_{i}N}{N}h^{ij}\frac{\partial_{j}N}{N}\phi_{0}^{*}\phi_{0}\right]
\end{align}
This solves Eq.~\eqref{con} with 
\begin{align}
\xi_{R}=-\frac{n}{4(n-1)}, \qquad \xi_{N\phi}=0, \qquad \xi_{NN}=\frac{n}{8}\,.
\end{align}
The extra freedom for $z=2$ arises  from the fact that
$\phi_{0}^{*}\left[R+(n-1)\frac{\nabla^{2}N}{N}-\frac{(n-1)(n+2)}{4}\frac{\partial_{i}N}{N}\frac{\partial^{i}N}{N}\right]\phi_{0}$
is Weyl invariant. This special invariant quantity is  available only for $z=2$.

Having constructed a classically Weyl invariant curved space action, we have that $\tilde W=W-W_{\text{c.t.}}=W+\Delta S$ is invariant under these local transformations:
\begin{equation}
\label{eq:tildeWinv}
\tilde W[e^{-z\sigma} N,e^{-2\sigma}h_{ij}, g^\alpha(e^{-\sigma}\mu)]=
\tilde W[ N,h_{ij}, g^\alpha(\mu)]
\end{equation}
We have suppressed the explicit dependence on space and time and have
assumed the only dependence on the renormalization scale $\mu$ is
implicitly through the couplings: using $\mu$-independence of bare
couplings, 
$g_0=\mu^{k\epsilon}g(\mu)Z_g(g(\mu))=(\lambda\mu)^{k\epsilon}g(\lambda\mu)Z_g(g(\lambda\mu))$
so that
$(\lambda^{-1}\mu)^{k\epsilon}g(\mu)Z_g(g(\mu))=\mu^{k\epsilon}g(\lambda\mu)Z_g(g(\lambda\mu))$. 

The generating functional $W$ is not invariant in the sense of
Eq.~\eqref{eq:tildeWinv}. The anomalous variation of $W$ arises purely
from the counter-terms: under an infinitesimal transformation,
\begin{align}
\Delta_\sigma
W&=W_{\text{c.t.}}[(1-z\sigma)N,(1-2\sigma)h_{ij},g^\alpha-\sigma
\:\mu\: dg^\alpha/d\mu]-W_{\text{c.t.}}[N,h_{ij},g^\alpha]\nonumber\\
\label{eq:Anom2}
&=\int\  dt\ d^{d}x\
N\sqrt{h}\left(\text{terms with derivatives on $N$, $h_{ij}$,
    $g^\alpha$ and  $\sigma$} \right)
\end{align}
does not vanish. Using Eqs.~\eqref{eq:TOgiven} and choosing $\sigma$
to be an infinitesimal local test function,  this reads
\begin{equation}
\label{eq:traceAnom}
z\langle T^0{}_0\rangle+\langle T^i{}_i\rangle-\beta^\alpha\langle [\mathcal{O}_\alpha]\rangle=\left(\text{terms with derivatives on $N$, $h_{ij}$,
    $g^\alpha$ and  $\sigma$} \right)\,.
\end{equation}
Evaluating at space and time independent coupling constants and on a
flat metric, so that the right hand side vanishes, we recognize this
as the trace anomaly for NRQFT.

Since the Weyl group is Abelian, consistency conditions follow from requiring that
\begin{equation}
\label{eq:conds}
\left[\Delta_{\sigma},\Delta_{\sigma^{\prime}}\right]W=0\,.
\end{equation}
These consistency conditions impose relations on the various anomaly
terms on the right hand side of Eq.~\eqref{eq:Anom2}. In the following sections we classify
all possible anomaly terms and derive the relations imposed by these
conditions.

\subsection{Dynamical exponent}
In the theory of critical phenomena the dynamical exponent $\zeta$
characterizes how a correlation length scales with time in time
dependent correlations. At the classical level (the gaussian fixed point) this just corresponds
to the anisotropic exponent $z$ introduced above. To understand the
connection between these we must retain explicitly the dependence on
the mass parameter(s) $m$ in Eqs.~\eqref{eq:SAnomScalingCurved}
and~\eqref{eq:tildeWinv}. We consider for simplicity the case of a
single mass parameter. In particular, we have
\begin{equation}
\label{eq:tildeWinv-mass}
\tilde W[e^{-z\sigma} N,e^{-2\sigma}h_{ij}, g^\alpha(e^{-\sigma}\mu) , m(e^{-\sigma}\mu)]=
\tilde W[ N,h_{ij}, g^\alpha(\mu),m(\mu)]\,.
\end{equation}

By dimensional analysis and translational and rotational invariance,
the correlator of fundamental fields is given by
\[
\langle \phi(\vec x, t)\phi(0,0)\rangle =\frac1{|\vec x|^{2\Delta}}F(\ln(m(\mu)|\vec
x|^z/t),\ln(\mu|\vec x|))\,,
\]
for some dimensionless function of two arguments, $F(x,y)$. This
function is further constrained by the renormalization group
equation. At a fixed point, $\beta^\alpha=0$, it takes the form
\[
\left(\mu\frac{\partial}{\partial\mu}+\gamma_m
  m\frac{\partial}{\partial m}+2\gamma\right) \langle \phi(\vec x,
t)\phi(0,0)\rangle =0\,,
\]
where $\gamma_m$ and $\gamma$ are the mass anomalous dimension and the
field anomalous dimension, respectively. These are generally
dimensionless functions
of the dimensionless coupling constants, $g^\alpha$, here evaluated at
their fixed point values, say, $g^\alpha_*$. It follows that 
\[
\langle \phi(\vec x, t)\phi(0,0)\rangle =\frac1{\mu_0^{2\gamma}|\vec
  x|^{2(\Delta+\gamma)}}f(m(\mu_0)\mu_0^{-\gamma_m}|\vec x|^{z-\gamma_m}/t)\,.
\]
Here $\mu_0$ is a reference renormalization point and $f$ is a
dimensionless function of one variable. This shows that at the fixed
point the fields scale with dimension $\Delta+\gamma$ and the
dynamical exponent is $\zeta=z-\gamma_m$. It is important to
understand that while $\zeta$ can be thought of as running along
flows, the exponent $z$ is fixed to its classical (gaussian fixed point) value.

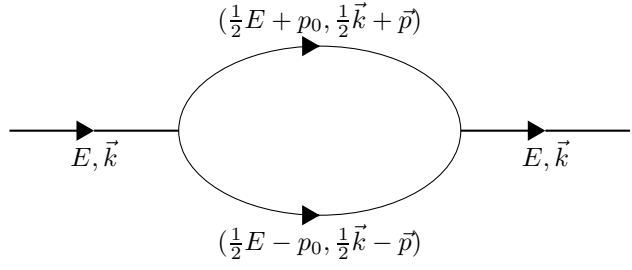
\begin{figure}[t]
\begin{center}
\begin{tikzpicture}[scale=0.75]
  \draw [thick, ->, >=triangle 60] (-3,0) --   (-1.5,0) node[anchor=north] {$E,\vec k$}  ;
  \draw [thick] (-1.5,0)   --   (0,0);
  \draw [thick, ->, >=triangle 60] (5,0)   --   (6.5,0)  node[anchor=north] {$E,\vec k$} ;
 \draw [thick] (6.5,0)   --   (8,0);
  \draw [thick, ->, >=triangle 60] (2.49,1.5)   --   (2.5,1.5);
 \draw [thick, ->, >=triangle 60] (2.49,-1.5)   --   (2.5,-1.5);
\draw (2.5,0) ellipse (2.5 and 1.5);
\node at (2.5,1.5) [anchor=south] {$(\frac12 E+p_0,\frac12 \vec k+\vec p)$};
\node at (2.5,-1.5) [anchor=north] {$(\frac12 E-p_0,\frac12 \vec k-\vec p)$};
\end{tikzpicture}
\end{center}
\caption{\label{fig:feynfig} Self energy correction to propagator at one loop }
\end{figure}

As an example consider the following Lagrangian for a $z=2$ theory  in $4+1$ dimensions:
\begin{equation}
\mathcal{L}= \left[iZ_m m\, Z_{\phi}\phi^*\overleftrightarrow{\partial_t}\phi-
  Z_{\phi}\vec\nabla\phi^* \vec\nabla\phi-\tfrac12Z_{g}g\mu^{\frac{\epsilon}{2}}\sqrt{Z_mm}\,Z_\phi^{3/2}|\phi|^{2}\left(\phi+\phi^{*}\right)\right]\,,
\end{equation}
The renormalization factors in dimensional regularization in $n+1$
dimensions, with $n=4-\epsilon$, have the following form:
\begin{equation}
  Z_{X}=1+\sum_{n=1}\frac{a^X_{n}}{\epsilon^{n}}\,,
\end{equation}
where the residues $a^X_n$ are functions of the renormalized coupling constant $g$. Independence of the bare parameters on the scale $\mu$ requires 
\begin{equation}
  0=\mu\frac{d}{d\mu}\left(Z_{g}g\mu^{\frac{\epsilon}{2}}\right)=\frac{\partial Z_{g}}{\partial g}\hat\beta g\mu^{\frac{\epsilon}{2}}+Z_{g}\hat\beta\mu^{\frac{\epsilon}{2}}+ \frac{\epsilon}{2}Z_{g}g\mu^{\frac{\epsilon}{2}}
\end{equation}
where $\hat\beta\equiv\mu dg/d\mu$ has $\hat\beta(g,\epsilon)=-\frac{\epsilon}2g+\beta(g)$, and
\begin{equation}
  0=\mu\frac{d}{d\mu}\left(Z_m m\right)=\frac{\partial Z_{m}}{\partial g}\hat \beta m+ \mu\frac{dm}{d\mu} Z_m\,.
\end{equation}
It follows that   
\begin{equation}
\gamma_{m}=\mu\frac{d\ln(m)}{d\mu}= \tfrac{1}{2}g\,\frac{da^m_1}{dg}\,.
\end{equation}

At one loop the self-energy correction to the propagator, represented
by the Feynman diagram in Fig.~\ref{fig:feynfig}, reads 
\begin{equation}
i \Sigma (E,\vec{k}) = -\tfrac12mg^{2}\mu^\epsilon \int \frac{dp_0}{2\pi}\frac{d^{n}p}{(2\pi)^n} \ D\left(\tfrac12E-p_{0},\tfrac12\vec{k}-\vec{p}\right)\ D\left(\tfrac12{E}+p_{0},\tfrac12\vec{k}+\vec{p}\right)
\end{equation}
where the propagator  is given by 
\begin{equation}
D(E,\vec{p}\:)= \frac{i}{\left(2mE-\vec p\,^{2}+i0^+\right)}\,.
\end{equation}
The integration over $p_{0}$ and then over $\vec p$ gives
\begin{equation}
\Sigma (E,\vec{k}\:) = \tfrac18g^{2}\mu^\epsilon  \int
\frac{d^{n}p}{(2\pi)^n}\ \frac{1}{\left(mE-\tfrac14{\vec k^{2}}-\vec
    p\,^{2}\right)}=-\frac{1}{\epsilon}\frac{g^2}{64\pi^2}(mE-\tfrac14\vec k\,^2)+\cdots\,,
\end{equation}
where the ellipses stand for finite terms. We read off
\[
Z_\phi-1 = \frac{g^2}{256\pi^2 \epsilon}\qquad \text{and}
\qquad Z_m-1=\frac{g^2}{256\pi^2 \epsilon}\,.
\]
Form which it follows that
\begin{equation}
\gamma_m=\frac{g^2}{256\pi^2}\,.
\end{equation}

\section{$d=2$, $z=2$ Non Relativistic theory}
\label{sec:2dimz2}
\subsection{Listing out terms} 
We first consider $2+1$ NRCFT with $z=2$. It is
convenient to catalogue the possible terms on the right hand side of
Eq.~\eqref{eq:Anom2} by the number of space and time derivatives
acting on the metric, the couplings and the transformation parameter
$\sigma$. Rotational invariance implies that space derivatives always
appear in contracted pairs. We must, in addition, insure the correct
dimensions. Table~\ref{tab:rotInvOps} summarizes the dimensions of the basic
rotationally invariant quantities; $R$ stands for the curvature scalar
constructed from the spatial metric~$h_{ij}$. Since $h_{ij}$ is the metric of a $2$
dimensional space, rotational invariants constructed from the Riemann  and  Ricci tensors can be
expressed in terms of $R$ only. 

\begin{table}[h]
    \centering
    \begin{tabular}{|c|c|c|c|}
    \hline
    Operators & $N$ & $g^{\alpha}$ & $R$\\
    \hline
    Length Dimension & $0$ & $0$ & $2$\\
    \hline
    Time Dimension & $1$ & $0$ & $0$\\
    \hline
        \end{tabular}
\caption{\label{tab:rotInvOps}Basic rotationally invariant operators and their dimensions.}
\end{table}

In order to match up the dimension of the Lagrangian, 
terms that only contain spatial derivatives must have exactly four
derivatives.  The derivatives can  act on the metric or on on the dimensionless
variation parameter $\sigma$. Hence we have following 2-spatial-derivatives
components:
\begin{gather}
\label{b1}\frac{\partial_{i}N}{N}\frac{\partial^{i}N}{N}\,,\quad
\frac{\partial_{i}N}{N}\partial^{i}g^{\alpha}\,,\quad
\partial_{i}g^{\alpha}\partial^{i}g^{\beta}\,,\quad
\frac{\nabla^{2}N}{N}\,,\quad
\nabla^{2}g^{\alpha}\,,\quad
R\\
\label{c1}\nabla^{2}\sigma\\
\label{d1}\partial_{i}\sigma \frac{\partial^{i}N}{N}\,,\quad
\partial_{i}\sigma \partial^{i}g^{\alpha}
\end{gather}
where we note that in the term $\frac{\partial_{i}N}{N}$ the
denominator serves to cancel off the time dimension of the numerator. 
To form a 4 derivative term out of above terms, we can 
(i)~choose two terms among \eqref{b1} with repetition
  allowed: there are $6^{2}-{}^6C_{2}=21$ such terms; (ii)~\eqref{c1} can combine with any of \eqref{b1}
  giving $6$ additional terms; and~(iii) we can choose one of \eqref{d1}  and choose
  another from \eqref{b1}, yielding an additional $2*6=12$ terms. 
Hence we will have $21+12+6=39$ terms with four space
derivatives. Terms with derivatives of $R$, such as
\[
\label{Rg}\partial_{i}R\partial^{i}g^{\alpha}\qquad\text{and}\qquad \partial_{i}R\frac{\partial^{i}N}{N}\,,
\]
are not independent. Integrating by parts, the term $\partial_{i}R\partial^{i}g^{\alpha}$ can written in terms of $R\nabla^{2}g^{\alpha}$ and
$R\partial_{i}\sigma\partial^{i}g^{\alpha}$, and the term
$R\nabla^{2}N$ can be expressed in terms of 
$\partial_{i}R\frac{\partial^{i}N}{N}$. The 39 four derivative terms,
which we call the $\nabla^4$ sector, appear on the right hand side of
\eqref{eq:Anom2} with dimensionless coefficients that are functions
of the couplings $g^\alpha$, and with a factor of $\sigma$ if the term
does not already contain one. Table~\ref{tab:XXXX} gives our notation
for the coefficients of these terms in Eq.~\eqref{eq:Anom2}.
\begin{table}[t]
    \centering
     \begin{tabular}{|c|c|c|c|c|c|c|c|c|c|}
        \hline
               \cellcolor{palegreen}$\nabla^4$ Sector & \cellcolor{lightyellow} $\partial_{i}N\partial^{i}N$  & \cellcolor{lightyellow} $\partial_{i}g^{\alpha}\partial^{i}g^{\beta}$& \cellcolor{lightyellow}$\partial_{i}N\partial^{i}g^{\alpha}$&\cellcolor{lightyellow} $\nabla^{2}N$&\cellcolor{lightyellow}$\nabla^{2}g^{\alpha}$&\cellcolor{lightyellow}$R$&\cellcolor{lightyellow}$\nabla^{2}\sigma$&\cellcolor{lightyellow}$\partial_{i}\sigma\partial^{i}N$&\cellcolor{lightyellow}$\partial_{i}\sigma\partial^{i}g^{\alpha}$ \\
        \hline
         \cellcolor{lightyellow} $\partial_{i}N\partial^{i}N$& \cellcolor{palegreen}$P_3, p_{3}$& $X_{\alpha\beta},x_{\alpha\beta}$ &$P_{1\alpha},\rho_{9\alpha}$&$P_4, p_{4}$&$Y_{\alpha},y_{\alpha}$&$Q,\chi_{4}$&$\chi_{3}$&$\rho_{11}$&$\rho_{8\alpha}$\\
        \hline
 \cellcolor{lightyellow} $\partial_{i}g^{\alpha}\partial^{i}g^{\beta}$ & \cellcolor{palegreen}$X_{\alpha\beta},x_{\alpha\beta}$& \cellcolor{palegreen}$X_{\alpha\beta\gamma\delta},x_{\alpha\beta\gamma\delta}$&$X_{\alpha\beta\gamma},x_{\alpha\beta\gamma}$&$X_{2\alpha\beta},x_{2\alpha\beta}$&$T_{2\alpha\beta\gamma},t_{2\alpha\beta\gamma}$&$Y_{5\alpha\beta},y_{5\alpha\beta}$&$a_{3\alpha\beta}$&$\rho_{1\alpha\beta}$&$t_{\alpha\beta\gamma}$\\
        \hline
         \cellcolor{lightyellow}$\partial_{i}N\partial^{i}g^{\alpha}$ &\cellcolor{palegreen}$P_{1\alpha},\rho_{9\alpha}$ &\cellcolor{palegreen}$X_{\alpha\beta\gamma},x_{\alpha\beta\gamma}$&\cellcolor{palegreen} $P_{5\alpha\beta},p_{5\alpha\beta}$ &$P_{25\alpha},\rho_{25\alpha}$& $P_{26\alpha\beta},\rho_{26\alpha\beta}$& $\chi_{\alpha}$& $\chi_{1\alpha}$& $\rho_{10\alpha}$& $x_{1\alpha\beta}$\\ 
         \hline
         \cellcolor{lightyellow} $\nabla^{2}N$&\cellcolor{palegreen}$P_4,p_{4}$&\cellcolor{palegreen}$X_{2\alpha\beta},x_{2\alpha\beta}$&\cellcolor{palegreen}$P_{25\alpha},\rho_{25\alpha}$&\cellcolor{palegreen}$P_{23},\rho_{23}$&$P_{24\alpha},\rho_{24\alpha}$&$H,c$\footnotemark &$h_{2}$&$\rho_{12}$&$\rho_{13\alpha}$\\
         \hline
         \cellcolor{lightyellow}$\nabla^{2}g^{\alpha}$&\cellcolor{palegreen}$Y_{\alpha},y_{\alpha}$&\cellcolor{palegreen}$T_{2\alpha\beta\gamma},t_{2\alpha\beta\gamma}$&\cellcolor{palegreen}$P_{26\alpha\beta},\rho_{26\alpha\beta}$&\cellcolor{palegreen}$P_{24\alpha},\rho_{24\alpha}$&\cellcolor{palegreen}$P_{22\alpha\beta},\rho_{22\alpha\beta}$&$A_{5\alpha},a_{5\alpha}$&$a_{4\alpha}$&$\rho_{7\alpha}$&$\rho_{21\alpha\beta}$
     \\
         \hline
         \cellcolor{lightyellow}$R$&\cellcolor{palegreen}$Q,\chi_{4}$&\cellcolor{palegreen}$Y_{5\alpha\beta}, y_{5\alpha\beta}$&\cellcolor{palegreen} $Q_{1\alpha},\chi_{\alpha}$&\cellcolor{palegreen}$H,c$ \footnotemark[1]\footnotetext[1]{$R\nabla^{2}N$ can be written as $\partial_{i}R\partial^{i}N$ by integration by parts, and it is for the operator $\partial_{i}R\partial^{i}N$    that  we      use    the coefficient~$c$.}&\cellcolor{palegreen}$A_{5\alpha}, a_{5\alpha}$&\cellcolor{palegreen}$A$, $a$&$n$ &$h_{1}$&$a_{7\alpha}$ \\
         \hline
         \cellcolor{lightyellow}$\nabla^{2}\sigma$&\cellcolor{palegreen}$\chi_{3}$&\cellcolor{palegreen}$a_{3\alpha\beta}$&\cellcolor{palegreen}$\chi_{1\alpha}$&\cellcolor{palegreen}$h_{2}$&\cellcolor{palegreen}$a_{4\alpha}$&\cellcolor{palegreen}$n$&\cellcolor{lightred}NA&\cellcolor{lightred}NA&\cellcolor{lightred}NA\\
         \hline 
         \cellcolor{lightyellow}$\partial_{i}\sigma\partial^{i}N$&\cellcolor{palegreen}$\rho_{11}$&\cellcolor{palegreen}$\rho_{1\alpha\beta}$&\cellcolor{palegreen}$\rho_{10\alpha}$&\cellcolor{palegreen}$\rho_{12}$&\cellcolor{palegreen}$\rho_{7\alpha}$&\cellcolor{palegreen}$h_{1}$&\cellcolor{lightred}NA&\cellcolor{lightred}NA&\cellcolor{lightred}NA\\
         \hline
         \cellcolor{lightyellow}$\partial_{i}\sigma\partial^{i}g^{\alpha}$&\cellcolor{palegreen}$\rho_{8\alpha}$&\cellcolor{palegreen}$t_{\alpha\beta\gamma}$&\cellcolor{palegreen}$x_{1\alpha\beta}$&\cellcolor{palegreen}$\rho_{13\alpha}$&\cellcolor{palegreen}$\rho_{21\alpha\beta}$&\cellcolor{palegreen}$a_{7\alpha}$&\cellcolor{lightred}NA&\cellcolor{lightred}NA&\cellcolor{lightred} NA\\ \hline
    \end{tabular}
    \caption{\label{tab:XXXX}  Summary of four spatial derivative terms
      that can enter the counterterm functional $W_{\text{c.t.}}$ or the
      anomaly on the  right hand side of  Eq.~\eqref{eq:Anom2}. The
      terms in $W_{\text{c.t.}}$  are the 
      products of the first six entries of the first column and the
      first six of the first  row, and their coefficients are the
      first of the 
      entries listed in the table (uppercase letters).  Those in the
      anomaly extend over the whole table; in the first $6\times6$
      block they correspond to the second entry (lowercase characters) and
      for those a factor of
      $\sigma$ is implicit. The red $NA$ labels
      denote terms that are second order in infinitesimal parameter
      $\sigma$, hence dropped. Latin indices are contracted with  the
      inverse metric $h^{ij}$ when repeated, eg, $\partial_{i}N\partial^{i}N=h^{ij}\partial_{i}N\partial_{j}N$.}
\end{table}

\begin{table}[h]
    \centering
    \begin{tabular}{|c|c|c|c|}
    \hline
    Operators & $K$ & $g^{\alpha}$ & $(K_{ij}-\frac{1}{2}Kh_{ij})$\\
    \hline
    Length Dimension & $0$ & $0$ & $0$\\
    \hline
    Time Dimension & $1$ & $0$ & $1$\\
    \hline
        \end{tabular}
\caption{\label{tab:Tops} Basic building blocks for operators in the $\partial_t^2$
  sector and their dimensions.}
\end{table}
Two time derivatives are required for the sector with pure time
derivatives, which we label $\partial_t^2$. The terms must still have
length dimension $-4$. The dimensions of the basic building blocks are
given  in
Tab.~\ref{tab:Tops}, where $K_{ij}=\frac12\partial_th_{ij}/N$ is the
extrinsic curvature of the $t=$constant hypersurfaces in the $N^i=0$
gauge and $K=h^{ij}K_{ij}$ (with $h^{ij}$ the inverse of the metric
$h_{ij}$). The combination $(K_{ij}-\frac{1}{2}Kh_{ij})$ is convenient
because it is Weyl invariant.  Hence, for the $\partial_t^2$ sector we
have the following basic one derivative terms:
\begin{gather}
\label{e1}K,\qquad\partial_{t}g^{\alpha}\\
\label{f1}\partial_{t}\sigma\\
\label{f2} K_{ij}-\tfrac{1}{2}Kh_{ij}
\end{gather}
The term $\partial_{t}N$ is not included in the list because it is not
covariant. The diffeomorphism invariant quantity is given by $\partial_{t}N-\Gamma^{0}_{00}N$ which vanishes identically $0$.

\begin{table}
    \centering
    \begin{tabular}{|c|c|c|c|c|}
        \hline
        \cellcolor{palegreen}$\partial_t^2$ Sector & \cellcolor{lightyellow} $K$  & \cellcolor{lightyellow} $\partial_{t}g^{\alpha}$& \cellcolor{lightyellow}$\partial_{t}\sigma$& \cellcolor{lightyellow} $K_{ij}-\tfrac{1}{2}Kh_{ij}$\\
        \hline
         \cellcolor{lightyellow} $K$& \cellcolor{palegreen}$ D,d$& $W_{\alpha}$, $w_{\alpha}$ &$f$& \cellcolor{lightred}NA\\
        \hline
 \cellcolor{lightyellow} $\partial_{t}g^{\alpha}$ & \cellcolor{palegreen}$W_{\alpha}$, $w_{\alpha}$& \cellcolor{palegreen}$X_{0\alpha\beta}$, $\chi_{0\alpha\beta}$&$b_{\alpha}$& \cellcolor{lightred}NA\\
        \hline
         \cellcolor{lightyellow}$\partial_{t}\sigma$ &\cellcolor{palegreen}$f$ &\cellcolor{palegreen}$b_{\alpha}$&\cellcolor{lightred} NA& \cellcolor{lightred}NA \\ \hline
         \cellcolor{lightyellow} $K_{ij}-\frac{1}{2}Kh_{ij}$& \cellcolor{lightred}NA& \cellcolor{lightred}NA& \cellcolor{lightred}NA &\cellcolor{palegreen}$E, e$\\
         \hline
    \end{tabular}
    \caption{\label{tab:TT}Summary of two time derivative terms
      that can enter the counterterm functional $W_{\text{c.t.}}$ or the
      anomaly on the  right hand side of  Eq.~\eqref{eq:Anom2}. The
      terms in $W_{\text{c.t.}}$  are the 
      products of the first, second, fourth entries of the first column and the
      first, second, fourth entry of the first  row , and their coefficients are the
      first of the 
      entries listed in the table (uppercase letters). Those in the
      anomaly extend over the whole table; in the first $2\times2$
      block they correspond to the second entry (lowercase characters) and
      for those a factor of
      $\sigma$ is implicit. The
      red NA labels denote terms that are either second order in
      infinitesimal parameter $\sigma$ or terms that are not rotationally
      invariant.} 
\end{table}

Possible anomaly terms are constructed from the $2^{2}-1=3$  products of
terms in \eqref{e1}; from 2 terms by combining \eqref{f1} and one from \eqref{e1}; and we can have \eqref{f2} contracted with itself. Thus in
total there are $3+2+1=6$ terms listed in Tab.~\ref{tab:TT} that also
gives the corresponding coefficients.

The sector with mixed derivatives has terms with one time and two
spatial derivatives. For this $\partial_t\nabla^2$  sector  we can
form terms by combining  one of \eqref{e1}  or \eqref{f1}
with one of  \eqref{b1}, \eqref{c1} or \eqref{d1}, excluding terms quadratic in $\sigma$. This gives
$3*9-3=24$ terms, as displayed with their coefficients in
Tab.~\ref{tab:TXX}. Finally, we have terms that are not constructed as products of
rotationally invariant quantities. Coefficient of those terms are
listed in the last row of~Tab.~\ref{tab:TXX}.

\begin{table}
    \centering
    \begin{tabular}{|c|c|c|c|c|c|c|c|c|c|}
      \hline
      \cellcolor{palegreen}$\partial_t\nabla^2$ Sector & \cellcolor{lightyellow} $\partial^{i}N\partial^{j}N$  & \cellcolor{lightyellow} $\partial^{i}g^{\alpha}\partial^{j}g^{\beta}$& \cellcolor{lightyellow}$\partial^{i}N\partial^{j}g^{\alpha}$&\cellcolor{lightyellow} $\nabla^{i}\nabla^{j}N$&\cellcolor{lightyellow}$\nabla^{i}\nabla^{j}g^{\alpha}$&\cellcolor{lightyellow}$R$&\cellcolor{lightyellow}$\nabla^{i}\nabla^{j}\sigma$&\cellcolor{lightyellow}$\partial^{i}\sigma\partial^{j}N$&\cellcolor{lightyellow}$\partial^{i}\sigma\partial^{j}g^{\alpha}$ \\
      \hline
      \cellcolor{lightyellow} K&\cellcolor{palegreen}$P$,
                            $\rho_{4}$&\cellcolor{palegreen}$X_{5\alpha\beta},x_{5\alpha\beta}$&\cellcolor{palegreen}$P_{\alpha}$,$\rho_{\alpha}$&\cellcolor{palegreen}$L,j$\footnote{$K\nabla^{2}N$
   can   be   written   as   $\partial_{i}K\partial^{i}N$   by   doing   integration by  parts,   and   it   is   for   this   operator   that   we   use   the   coefficient $j$. }&\cellcolor{palegreen}$P_{3\alpha},b_{8\alpha}$&\cellcolor{palegreen}$B,b
$&\cellcolor{palegreen}$m$&\cellcolor{palegreen}$l_{1}$&\cellcolor{palegreen}$b_{7\alpha}$\\
      \hline
      \cellcolor{lightyellow}$\partial_{t}g^{\alpha}$&\cellcolor{palegreen}$X_{\alpha},\rho_{6\alpha}$&\cellcolor{palegreen}$X_{3\alpha\beta\gamma},x_{3\alpha\beta\gamma}$&\cellcolor{palegreen}$P_{4\alpha\beta},p_{4\alpha\beta}$&\cellcolor{palegreen}$B_{6\alpha},b_{6\alpha}$&\cellcolor{palegreen}$X_{4\alpha\beta},x_{4\alpha\beta}$&\cellcolor{palegreen}$B_{5\alpha},b_{5\alpha}$&\cellcolor{palegreen}$B_{9\alpha},b_{9\alpha}$&\cellcolor{palegreen}$\rho_{5\alpha}$&\cellcolor{palegreen}$x_{6\alpha\beta}$\\
      \hline
      \cellcolor{lightyellow}$\partial_{t}\sigma$&\cellcolor{palegreen}$\rho_{3}$&\cellcolor{palegreen}$b_{3\alpha\beta}$&\cellcolor{palegreen}$\rho_{1\alpha}$&\cellcolor{palegreen}$l_{2}$&\cellcolor{palegreen}$b_{4\alpha}$&\cellcolor{palegreen}$k$&\cellcolor{lightred}NA&\cellcolor{lightred}NA&\cellcolor{lightred}NA\\
        
        \hline
        \cellcolor{lightyellow}$K_{ij}-\frac{1}{2}Kh_{ij}$ &\cellcolor{palegreen}$F_1,f_{1}$ &\cellcolor{palegreen}$F_{2\alpha\beta}, f_{2\alpha\beta}$&\cellcolor{palegreen}$F_{3\alpha}, f_{3\alpha}$&\cellcolor{palegreen}$F_4, f_{4}$&\cellcolor{palegreen}$F_{5\alpha}, f_{5\alpha}$&\cellcolor{lavender}NA &\cellcolor{palegreen} $f_6$&\cellcolor{palegreen}$f_{7}$&\cellcolor{palegreen}$f_{8\alpha}$\\
        \hline
         \end{tabular}
         \caption{\label{tab:TXX}Summary of one-time, two-space derivative terms
           that can enter the counterterm functional $W_{\text{c.t.}}$ or the
           anomaly on the  right hand side of  Eq.~\eqref{eq:Anom2}. The
           terms in $W_{\text{c.t.}}$  are the 
           products of the entries that have no explicit $\sigma$
           factor, and their coefficients are the  first of the 
           entries listed in the table (uppercase letters).  Those in the
           anomaly extend over the whole table; terms without explicit
           $\sigma$ have coefficients that correspond to the second
           entry (lowercase characters) and  for those a factor of
           $\sigma$ must be included. Latin indices are contracted with
           the spatial metric  as necessary to make  the product of the first column and
           first row entries rotationally invariant; for example, $\rho_{4}$
           denotes the coefficient of  $K\partial_{i}N\partial^{i}N$. For
           last entry in the first column,   indices are contracted
           with those in the terms in first row. The red NA labels
           denote terms that are second order in infinitesimal
           parameter $\sigma$, hence dropped. The blue NA
           one denotes a term that is identically $0$ since
           $K_{ij}-\tfrac{1}{2}Kh_{ij}$ vanishes upon contraction via
           $h^{ij}$.} 
\end{table}

\subsection{Using counter-terms}
\label{sec:CTs}
One can similarly list all possible terms in $W_{\rm{c.t.}}$. The requirements imposed by
dimensional analysis and rotational invariance are as before, the only difference being
that these terms are built from the metric and the couplings but not the parameter of the
Weyl transformation $\sigma$. Therefore the list of possible counterterms is obtained from
the one for anomalies by replacing $\sigma\to 1$.  Tables.~\ref{tab:XXXX}, \ref{tab:TT} and~\ref{tab:TXX}
give, as uppercase letters, our notation for the coefficients of
these operators in $W_{\rm{c.t.}}$.

The counterterms in $W_{\text{c.t.}}$ are not completely fixed by
requiring finiteness of the generating functional. The ambiguity
consists of the freedom to include arbitrary finite contributions to
each term. This freedom to add finite counter-terms does not affect
the consistency conditions but does change the value of the individual
terms related by them.  We can use this freedom to set some 
anomalies to zero, simplifying the analysis of the consequences of the
Weyl consistency conditions.
In particular, in searching for an $a$-theorem we can use this freedom to simplify
the consistency conditions. It may be possible to show then that there exist some class of
subtraction schemes  for which there exists a possible candidate for an $a$-theorem, but
 a general, counter-term and scheme independent statement may not be possible.

\newpage
To illustrate this, consider the variation of the $K^2$ and $K\partial_{t}g^{\alpha}$  terms in $W_{\text{c.t.}}$:
\[
\Delta_\sigma \int dt\,d^2x \,N\sqrt{h}\left(D\,K^2\right)=\int
dt\,d^2x \,N\sqrt{h}\left(-4\frac1{N}\partial_t\sigma 
  DK-\sigma\beta^\alpha\partial_\alpha  D\, K^2\right)\,,
\]
\vspace{-35pt}
\begin{multline*}
\Delta_\sigma \int dt\,d^2x \,N\sqrt{h}\left( W_{\alpha}\,K\partial_{t}g^{\alpha}\right)=\int
dt\,d^2x \,N\sqrt{h} \, \bigg(\!-\sigma\left[\beta^\alpha\partial_\alpha  W_{\gamma}+
                       W_{\alpha}\partial_{\gamma}\beta^{\alpha}\right] \,
                       K\partial_{t}g^{\gamma} \\
 -\frac{1}{N}\partial_t\sigma\ \beta^{\alpha}\  
  W_{\alpha}K-2\frac{1}{N}\partial_t\sigma\   W_{\alpha}\partial_{t}g^{\alpha}\bigg)
\end{multline*}
Inspecting Tabs.~\ref{tab:XXXX}, \ref{tab:TT} and~\ref{tab:TXX} we see that the $f$
anomaly gets contributions only from these variations, so that the change in $f$
induced by finite changes in the counterterms is given by
\begin{equation}
\label{eq:example-f-ambig}
\delta f=-4\,   D\ -\beta^{\alpha}\,   W_{\alpha}\,.
\end{equation}
With a slight abuse of notation we have denoted here the arbitrary, finite, additive
change to the coefficients of counterterms by the same symbol we have used for the
counterterm coefficients themselves. From Eq.~\eqref{eq:example-f-ambig} we see that  one can always choose $D$ so as to set
$f$ arbitrarily, and it is often convenient to set $f=0$.  For a second
example consider the $R^2$ anomaly, $a$. A similar computation gives
\begin{equation}
\delta a =  -\beta^{\alpha}\partial_{\alpha}\,  A
\end{equation}
In this case we may solve this equation so as to
set $a=0$ only if $a=0$ at fixed points, where $\beta^\alpha=0$. As
we will see below, the Weyl consistency conditions constrain some
anomalies to vanish at fixed points. 

\begin{table}
\centering
\begin{tabular}{|c||c|}
\hline
Sector& Trivial Anomalies\\
\hline
\hline
$\partial_t^2$ & $f$, $b_\alpha$\\
\hline
$\nabla^2\partial_t$ & $(\rho_{3},l_1)$, $ x_{6\alpha\beta}$, $
                       \rho_{5\alpha}$, $ b_{3\alpha\beta}$,$b_{4\alpha}$, $ b_{9\alpha}$, $ (k, m, l_{2})$, $(
                     b_{7\alpha}, \rho_{1\alpha})$, $ f_6$, $f_7$, $f_{8\alpha}$ \\
\hline
$\nabla^4$ &$ \chi_3$, $ \rho_{11}$, $ (\rho_{10\alpha}, \rho_{13\alpha}, \rho_{8\alpha})$, $ a_{3\alpha\beta}$, $\rho_{1\alpha\beta}$, $t_{\alpha\beta\gamma}$,  $ \chi_{1\alpha}$,  $ x_{1\alpha\beta}$, $ h_2$, $ \rho_{12}$, $ a_{4\alpha}$,  $\rho_{7\alpha} $, $\rho_{21\alpha\beta}$, $n$, $ h_1$,  $ a_{7\alpha}$ \\
\hline
\end{tabular}
\caption{\label{tab:triv} Trivial anomalies for each sector. Finite
  ambiguities in counter-terms give sufficient freedom to set all
  these anomalies arbitrarily; setting them to zero is often
  convenient. For anomalies grouped within parenthesis, all but one of them  can be
  set arbitrarily.}
\end{table}

We give in App.~\ref{app:anom-ambig} the complete set of ambiguities
for models with $z=2$ in $d=2$ spatial dimensions. Terms in the
effective actions whose coefficients can be varied at will are not
properly anomalies, since the coefficients can be set to
zero.  With a slight abuse of language they are commonly
  referred to as {\it trivial anomalies} and we adopt this terminology
  here. Table~\ref{tab:triv} summarizes the trivial anomalies found
in each sector.

\subsection{Consistency conditions and vanishing anomalies}
In computing the consistency condition \eqref{eq:conds} one finds a functional that is a
combination of linearly independent ``operators'' (combinations of $\sigma$,
$\gamma_{\mu\nu}$ and $g^\alpha$), each with a coefficient that is a linear combination of
the coefficients in Tabs.~\ref{tab:XXXX}, \ref{tab:TT} and~\ref{tab:TXX} and their
derivatives. Thus the consistency conditions can be expressed as a set of equations among
these coefficients and their derivatives. The full set of consistency conditions for
$d=2$, $z=2$ are listed in App.~\ref{app:conditions22}. On the left of each condition we
have listed the operator the condition arises from. We have verified that these conditions
reduce to the ones computed in Ref.~\cite{Baggio:2011ha} at fixed points. In the
$\partial_t\nabla^2$ sector the consistency conditions, Eqs.~\eqref{eqs:TXXconditions},  are given  for
arbitrary $z$, while for the  $\partial_t^2$ and $\nabla^4$   sectors, Eqs.~\eqref{eqs:TTconditions} and~\eqref{eqs:XXXXconditions},
respectively, the value  $z=2$ has been used.

\begin{table}
\centering
\begin{tabular}{|c||c||c|}
\hline
Sector & Vanishing Anomalies & Conditionally\\ & & Vanishing Anomalies\\
\hline
\hline
$\partial_t^2$ & $d$ & $w_{\alpha}$\\
\hline
$\nabla^2\partial_t$ & $f_4$, $f_1$, $ \rho_{4}$, $ b_{7\alpha}$ & $ b_{6\alpha}+\rho_{6\alpha}$, $b_{5\alpha}-b_{6\alpha}$, $b_{7\alpha}-\rho_{1\alpha}$ \\ &  $b$, $j$, $2\rho_{3}-l_{1}+2l_{2}$, $k+m-l_{2}$ &  $ x_{5\alpha\beta}$, $f_{3\alpha}$, $b_{8\alpha}$
 \\
\hline
$\nabla^4$ &$ \chi_4-p_4$, $ 2p_3+p_4$, $ c-\chi_4$, $h_1+2h_2+2\chi_3-c-\rho_{12}$ & $
             x_{\alpha\beta}+x_{2\alpha\beta}$, $ \rho_{13\alpha}$ \\ & $ 2a+c$, $ p_4+2\rho_{23}$, $ 2\rho_{23}+c$ & $
             y_{5\alpha\beta}-x_{2\alpha\beta}$, 
 $ y_{\alpha}+\rho_{24\alpha}$, $a_{5\alpha}-\rho_{24\alpha}$, $\rho_{25\alpha}+\rho_{9\alpha}$  \\
\hline
\end{tabular}
\caption{\label{tab:vanishing} Vanishing anomalies for each
  sector. The Weyl consistency conditions imply these anomalies, or
  combination of anomalies, vanish at fixed points (where
  $\beta^\alpha=0$). An anomaly is conditionally vanishing if it is vanishing only for a particular choice of counterterms.}
\end{table}

At fixed points the consistency conditions imply some anomalies
  vanish. These are known as {\it vanishing anomalies}. For example, setting
  $\beta^\alpha=0$ in Eq.~\eqref{dtK} gives $d=0$.  Table~\ref{tab:vanishing} summarizes
  the vanishing anomalies found in each sector. The table also shows {\it conditionally
  vanishing anomalies.} These are vanishing anomalies but only for a specific choice of
  counterterms. For example, setting $\beta^\alpha=0$ in Eq.~\eqref{dt2g} gives
  $-2w_\alpha+b_\gamma\partial_\alpha\beta^\gamma=0$, and
  Eq.~\eqref{eq:balpha-ambig} shows that we can choose the counterterm $W_\alpha$ to set
  $b_\alpha=0$.

As explained above, some vanishing anomalies can be set to zero. For
example, from Tab.~\ref{tab:vanishing} we see that $d$ is a vanishing anomaly, and then Eq.~\eqref{eq:d-ambig}
informs us that one may choose $D$ to enforce $d=0$. We note, however,
that by Eqs.~\eqref{eq:f-ambig} and~\eqref{eq:d-ambig} one may either choose
$f$ or $d$ to vanish, but not both.

\subsection{Applications}
\label{subsec:apps}
While there are many avenues for analysis in light of the relations
imposed by Weyl consistency conditions on the anomalies, we
concentrate on finding candidates for a C-theorem. We search for a
combination of anomalies, $C$, a local function in the space of
dimensionless coupling constants that flows monotonically,
$\mu dC/d\mu\ge0$. We try to establish this by judiciously setting
some anomalies to zero by the freedom explained above and looking for a relation
of the form
\[
\beta^\alpha\partial_\alpha C = -
\mathcal{H}_{\alpha\gamma}\beta^\alpha\beta^\gamma \,.
\]

Our first three candidates arise from the $\nabla^4$ sector. Consider Eq.~\eqref{eq:del2R}, here reproduced:
\[
-a_{5\alpha}\beta^{\alpha}+4a+2c+\beta^{\alpha}\partial_{\alpha}n=0
\]
The combination $2a+c$ is a vanishing anomaly. One may then use
\eqref{eq:a-ambig} and \eqref{eq:c-ambig} to set $2a+c=0$.
Equation~\eqref{eq:a4-ambig} shows $a_{4\alpha}$ is a trivial anomaly
and one may set $a_{4\alpha}=0$. Combining with Eq.~\eqref{eq:del2sigdel2g} we have
\[
\beta^{\alpha}\partial_{\alpha}n=\rho_{22\alpha\gamma}\beta^{\alpha}\beta^{\gamma}+\rho_{24\alpha}\beta^{\alpha}
\]
Similarly, Eq.~\eqref{eq:del2N} shows $2\rho_{23}+c$ is a vanishing
anomaly and using \eqref{eq:rho23-ambig} we may set
$2\rho_{23}+c=0$. We then have from Eq.~\eqref{eq:del2N} again that 
\[
\beta^{\gamma}\partial_{\gamma}h_{2}=\beta^{\gamma}\rho_{24\gamma}
\]
The difference of these equations then gives us our first candidate
for a C-theorem, with $C=n-h_2$:
\begin{equation}
\label{Ccand1}
\beta^{\alpha}\partial_{\alpha}(n-h_2)=\rho_{22\alpha\gamma}\beta^{\alpha}\beta^{\gamma}\,.
\end{equation}

A second candidate can be found  as follows. Eq.~\eqref{eq:dsdNdN} shows
$\chi_4-p_4$ is a vanishing anomaly. Then $Q-P_4$ can be chosen so
that $\chi_4-p_4=0$; see Eqs.~\eqref{eq:chi-ambig} and~\eqref{eq:p4-ambig}. Using Eq.~\eqref{eq:dsdNd2g}
with $\rho_{7\alpha}=0$ as it is a trivial anomaly, we obtain
\[
-\beta^\alpha\partial_\alpha\chi_3=\tfrac14\rho_{26\alpha\gamma}\beta^{\alpha}\beta^{\gamma}+\rho_{24\alpha}\beta^{\alpha}
\]
It follows that
\begin{equation}
\label{Ccand2}
\beta^\alpha\partial_\alpha\left(n+\chi_3\right)=(\rho_{22\alpha\gamma}-\tfrac14 \rho_{26\alpha\gamma})\beta^{\alpha}\beta^{\gamma}
\end{equation}

Combining Eqs.~\eqref{eq:d2sdgdN},~\eqref{eq:dsdNd2N}
and~\eqref{eq:dsRdN} while setting 
$\chi_{1\alpha}=0$, $p_4+2\rho_{23}=0$ and $c-\chi_4=0$ gives what
appears to be yet another candidtae in the $\nabla^4$ sector:
\begin{equation}
\label{Ccand3}
\beta^\alpha\partial_\alpha\left(c+\rho_{12}-h_1\right)=-\tfrac12 \rho_{26\alpha\gamma}\beta^{\alpha}\beta^{\gamma}
\end{equation}
However, setting the trivial anomalies $\rho_{1\alpha}$ and $\chi_{1\alpha}$ to   zero, Eq.~\eqref{eq:dsd2sdN} gives
\[
h_2+\chi_3=\tfrac12(c+\rho_{12}-h_1)
\]
which shows that the candidates given by
eq~\eqref{Ccand1},\eqref{Ccand2},\eqref{Ccand3} are not linearly independent in the
scheme with $2a+c=2\rho_{23}+c=\chi_4-c=\chi_4-p_4=p_4+\rho_{23}=0$ and $a_{4\alpha}=\rho_{1\alpha}=\rho_{7\alpha}=\chi_{1\alpha}=0$.

We find one candidate for a C-theorem in the $\partial_t^2$
sector. Equation~\eqref{dtK} shows $d$ is a vanishing anomaly and use
Eqs.~\eqref{eq:d-ambig} and~\eqref{eq:balpha-ambig} to set
$d=b_\alpha=0$. Combining \eqref{dtK}  and \eqref{dt2g} gives
\begin{equation}
\label{Ccand4}
\beta^\alpha\partial_\alpha f=-\chi_{0\alpha\gamma}\beta^{\alpha}\beta^{\gamma}\,.
\end{equation}

In the $\partial_t\nabla^2$-sector we find the following candidates for a $C$-theorem:
\begin{align}
\label{CcandXXT1}
\beta^{\alpha}\partial_{\alpha}m&=-\tfrac12x_{4\alpha\gamma}\beta^{\alpha}\beta^{\gamma}
\\
\label{CcandXXT2}
\beta^{\alpha}\partial_{\alpha} l_1&=-\tfrac12p_{4\gamma\alpha}\beta^{\gamma}\beta^{\alpha}
\\
\label{CcandXXT3}
\beta^{\alpha}\partial_{\alpha}\left(\rho_3+l_2\right)&=-\tfrac1{2z}p_{4\alpha\gamma}\beta^{\gamma}\beta^{\alpha}
\\
\label{CcandXXT4}
\beta^{\alpha}\partial_{\alpha}\left(f_6+\frac{z}{2}f_7-\beta^{\gamma}f_{5\gamma}\right)&=\beta^{\alpha}\beta^{\gamma}\left(f_{2\alpha\gamma}-\partial_{\alpha}f_{5\gamma}\right)
\end{align}
We have kept the explicit dependence on $z$ in these equations. As we will see below  the
$\partial_t\nabla^2$-sector is special in that the Weyl anomalies and the relations  from
consistency conditions hold for arbitrary $z$. Hence, the $C$-candidates in this sector  are
particularly interesting since they are  candidates  for  any $z$. To derive \eqref{CcandXXT1} we
have used that $j$ and $b$ are vanishing anomalies, as evident from Eqs.~\eqref{w1} and~\eqref{w2},
and used $B$  and $L$ to set $b=j=0$ in Eq.~\eqref{w2} and $P_{3\alpha}$ to set $b_{4\alpha}=0$ in
Eq.~\eqref{w0}.  For~\eqref{CcandXXT2} we used $j=0$ in Eq.~\eqref{u0} and \eqref{u1}, deduce that
$\rho_{4}$ is a vanishing anomaly and use $P$ to set $\rho_{4}=0$ in Eq.~\eqref{u1} and $P_{\alpha}$
to set $\rho_{1\alpha}=0$ in Eq.~\eqref{u0}.  For~\eqref{CcandXXT3},  we set $j=\rho_4=0$ as before
and in addition  we set $\rho_{5\alpha}=0$ using $X_{\alpha}$ in \eqref{q2}, and use
Eqs.~\eqref{w1}, \eqref{q2} and~\eqref{q3}.  In the scheme, $j=\rho_{1\alpha}=0$, Eq~\eqref{so0} implies that the candidates given by~\eqref{CcandXXT3} and~\eqref{CcandXXT2} are linearly dependent. Last but not the least, \eqref{CcandXXT4} is derived
from Eqs.~\eqref{so1}--\eqref{so3} by  using
$F_{3\alpha}$ to set $f_{8\alpha}=0$ and setting to
zero the vanishing anomalies $f_1$ and $f_4$ using $F_1$ and $F_4$. 

Two comments are in order. First, we have not established any $C$-theorem. To do so would
require showing that the two index symmetric tensor appearing on at least one of the right
hand side of Eqs.~\eqref{Ccand1}--\eqref{Ccand4} is positive definite, so that it acts as
a metric in the space of flows. In addition, the interpretation of $C$ as
  counting degrees of freedom is better supported if it is a monotonic function of
  the number of degrees of freedom at a gaussian fixed point. And second, we do not expect a positive definite metric
can be found in generality, since cyclic flows are known to appear in NR quantum
systems. Cyclic flows appear in relativistic systems too, but they differ from NR ones in
that there is scaling symmetry all along the cyclic flows and, in fact, the $C$ quantity
is constant along the cyclic flow \cite{Fortin:2012hn}. Investigating the conditions under
which a theory gives positive definite metric(s) in the space of flows is beyond the scope
of this work; we hope to return to this problem in the future.

\section{Generalisation to arbitrary $z$ value}
\label{sec:allz}
In this section, we will explore the possibility to generalize the
work for arbitrary $z$ value.  It is clear that the formalism fails
for non-integer values of $z$ since in that case, we can not make up for
dimensions with regular analytic functions of curvature and coupling
constants. This is because the quantities constructed out of geometry
and coupling constants always have integer length and time
dimension. Furthermore, in a Lagrangian formulation a non-integer $z$ requires
non-analyticity of Lagrangian. So we begin by recalling under what
conditions a Lagrangian with local interactions allows for integer $z$
values. 

Consider first the case of $d=2$ at arbitrary $z$ value. In
constructing $\Delta W_{\text{c.t.}}$, rotational invariance implies even number of spatial
derivatives, say $2n$. Along with $m$ time derivatives, we must have 
\begin{equation*}
mz+2n=z+2\,.
\end{equation*}
We look for solutions with integer values for  $m$ and $n$. For $m=1$ we must have $n=1$
and this  satisfies the equation for any $z$. Else, for  $m\neq1$ we have
\begin{equation*}
z=\frac{2(1-n)}{(m-1)}\,.
\end{equation*}
For $z>0$ we must have either $m=0$ with $n>1$ or $n=0$ with $m>1$.  For $m=0$ solutions
exist only if $z=2k$ is even, with
$2n=2(1+k)$ spatial derivatives. On the other hand, with $n=0$, we have
solutions for  $z=2/k$, with $m=k+2$ time
derivatives. 
To summarize, for $z>0$ we can classify the counterterms
by sector as follows:
\begin{itemize}
\item There is a pure $\nabla^{2}$ sector for $z=2k$, $k\in\mathbf{Z}$. It has precisely
  $2(k+1)$ spatial derivatives. We have discussed
  in detail
  the case $k=1$. Higher values of $k$ can be similarly
  analyzed, but it it involves an ever increasing number of terms as
  $z$ increases.   
\item There is a pure $\partial_t$ sector for $z=2/k$, $k\in\mathbf{Z}$,  with $k+1$ time
  derivatives.  We have analyzed the $k=1$ case. Higher values of $k$ can be similarly
  analyzed, but it involves an ever increasing number of terms as
  $z$ decreases.   
\item There is a  $\partial_t\nabla^2$ sector for  arbitrary 
  $z$.  It has 1-time and 2-spatal derivatives regardless of $z$. 
Therefore, the classification of anomalies and counterterms is
exactly as in the $z=2$ case, and the consistency conditions and
derived $C$-candidates are modified by factors of
$z/2$ relative the $z=2$ case. 
\end{itemize}

\section{A candidate for a $C$-theorem in $d+1$D  }
\label{sec:alld}
In relativistic $2n$-dimensional QFT the quantity that is believed to
satisfy a $C$-theorem is associated with the Euler anomaly, that is,
it is the coefficient of the Euler density $E_{2n}$ in the conformal
anomaly \cite{Grinstein:2013cka}.\footnote{There is no known local
  $C$-function candidate for odd-dimensional relativistic field
  theory. Jafferis has proposed a non-local $F$-function for 3D
  relativistic theories that shares the monotonicity properties of a
  $C$-function\cite{Jafferis:2010un}} It would seem natural to seek for
analogous candidates in non-relativistic theories. The obvious analog
involves the Euler density for the spatial sections $t=$~constant; by
dimensional analysis and scaling it should be constructed out of
$z+d=2n$ spatial derivatives acting on the metric $h_{ij}$. However,
for a $d$-dimensional metric the Euler density $E_{2n}$ with
$2n-d=z>0$ vanishes. Hence, we are led to consider an anomaly of the
form $XE_d$, that is the Euler density computed on the spatial
sections $t = \text{constant}$ times some quantity $X$ with the
correct dimensions, $[X]=z$. This construction is only valid for even spatial dimension, $d=2n$. The most natural candidate for $X$ is $K$: it is the only choice if $z$ is odd. If $z$ is even it can be
constructed out of spatial derivatives. For example, if $z=dk=2nk$ for
some integers $k$ and $n$, one may take $X=(E_{d})^{k}$.

The variation of the Euler density yields the Lovelock tensor \cite{lovelock},   $H_{ij}$, a
symmetric 2-index tensor that  satisfies 
\begin{equation*}
\nabla_{i}H^{ij}=0
\end{equation*}
In looking for a candidate $C$-theorem we consider a set of operators
that close under Weyl-consistency conditions, starting from
$XE_d$. Since $\delta_\sigma(\sqrt{h} E_d)=\sqrt{h}H^{ij}\nabla_i\partial_j\sigma$, and 
$[XH^{ij}]=z+d-2$,  we are led to include terms with the Lovelock tensor and two
spatial derivatives. In order to compute the consequences of the Weyl consistency conditions
we assume
\begin{equation}
\delta X = z\sigma X + \cdots
\end{equation}
where the ellipses denote terms that depend on derivatives of $\sigma$
and are therefore independent of $X$. Consider therefore  a subset of terms in the anomaly that appear in the consistency conditions that lead to a potential $C$-theorem:
\begin{multline}
\Delta_{\sigma} W= \int d^{d}x dt\ N\ \sqrt{h} \left[
  \sigma\left\{aXE_{d} +
 b XH^{ij}R_{ij}
+   \chi_4 XH^{ij}\frac{\partial_{i}N}{N}\frac{\partial_{j}N}{N}
+\chi_{\alpha} XH^{ij}\frac{\partial_{i}N}{N}\partial_{j}g^{\alpha}\right.\right.\\
\left.\left. 
+y_{5\alpha\beta}XH^{ij}\partial_{i}g^{\alpha}\partial_{j}g^{\beta}
+cH^{ij}\partial_{i}X\frac{\partial_{i}N}{N}
+a_{5\alpha} H^{ij}\partial_{i}X\partial_{j}g^{\alpha}
\right\}
\right.\\\left.
+\partial_i\sigma\left\{ n \partial_{j}XH^{ij} 
+ h_1 \frac{\partial_{j}N}{N}H^{ij}X
+ a_{7\alpha} \partial_{j}g^{\alpha}H^{ij}X
\right\}\right]
\end{multline}

Correspondingly there are metric and coupling-constant dependent
counter-terms with coefficients denoted by uppercase symbols: 
\begin{multline}
 W_{\text{c.t.}}= \int d^{d}x dt\ N\ \sqrt{h} \left[
  AXE_{d} 
+ B XH^{ij}R_{ij}+
   X_4 XH^{ij}\frac{\partial_{i}N}{N}\frac{\partial_{j}N}{N}
+X_{\alpha} XH^{ij}\frac{\partial_{i}N}{N}\partial_{j}g^{\alpha}\right.\\
\left.+Y_{5\alpha\beta}XH^{ij}\partial_{i}g^{\alpha}\partial_{j}g^{\beta}
+CH^{ij}\partial_{i}X\frac{\partial_{i}N}{N}
+A_{5\alpha} H^{ij}\partial_{i}X\partial_{j}g^{\alpha}
\right]
\end{multline}
Freedom to choose finite parts of counter-terms leads to ambiguities in
the anomaly coefficients as follows:
\begin{subequations}
\label{gen-ambig}
\begin{align}
\label{gen-ambig1}\delta a =&-\beta^{\alpha}\partial_{\alpha}A\\
\label{gen-ambig2}\delta \chi_{4} =&-\beta^{\alpha}\partial_{\alpha}X_{4}\\
\label{gen-ambig3}\delta \chi_{\alpha} =&-\beta^{\gamma}\partial_{\gamma}X_{\alpha}-X_{\gamma}\partial_{\alpha}\beta^{\gamma}\\
\label{gen-ambig4}\delta y_{5\alpha\beta} =& -\beta^{\gamma}\partial_{\gamma}Y_{5\alpha\beta}-Y_{5\gamma\beta}\partial_{\alpha}\beta^{\gamma}-Y_{5\alpha\gamma}\partial_{\beta}\beta^{\gamma}\\
\label{gen-ambig5}\delta c =& -\beta^{\alpha}\partial_{\alpha}C\\
\label{gen-ambig6}\delta b =& -\beta^{\alpha}\partial_{\alpha}B\\
\label{gen-ambig7}\delta a_{5\alpha} =& -\beta^{\gamma}\partial_{\gamma}A_{5\alpha}-A_{5\gamma}\partial_{\alpha}\beta^{\gamma}\\
\label{gen-ambig8}\delta n =& -A -(d-2)B -Cz -\beta^{\alpha}A_{5\alpha}\\
\label{gen-ambig9}\delta h_{1}=& -2zX_{4}-\beta^{\alpha}X_{\alpha}+ Cz -A -(d-2)B\\
\label{gen-ambig10}\delta a_{7\alpha} =& -\partial_{\alpha}\left(A+(d-2)B\right) -zX_{\alpha}-2\beta^{\gamma}Y_{5\gamma\alpha}+zA_{5\alpha}
\end{align}
\end{subequations}

In addition to the Euler density, $E_d$, there are several independent scalars one can construct out
of $d$ derivatives of the metric in $d$ dimensions (except for $d=2$, for which the only
2-derivative invariant is the Ricci scalar and hence $E_d\propto R$). $E_d$ is special in that it is
the only quantity that gives just the Lovelock tensor under an infinitesimal Weyl trasformation,
$\delta_\sigma (\sqrt{h}E_d)= \sqrt{h}H^{ij}\nabla_i\partial_j\sigma$.  In general some other
$d$-derivative invariant\footnote{Weyl variations of $d$-derivative scalars constructed from less
  than $d/2$ powers of the Riemann tensor do not contribute to the consistency condition we are considering.} $\mathcal{E}$ constructed out of $d/2$ powers of the Riemann tensor
will instead give
$\delta_\sigma(\sqrt{h}\mathcal{E})=\sqrt{h}\mathcal{H}^{ij}\nabla_i\partial_j\sigma$ where
$\mathcal{H}^{ij}\ne0$ is not divergence-less, $\nabla_i\mathcal{H}^{ij}\ne0$.  We have given an
example of such a term above, $H^{ij}R_{ij}$, both in the anomaly and among the counter-terms. Given
a basis of $d$-derivative operators $\mathcal{E}$ and $d-2$ derivative 2-index symmetric tensors
$\mathcal{H}^{ij}$ one can derive Weyl consistency conditions by demanding that the coefficients of
each linearly independent operator in $[\Delta_\sigma,\Delta_{\sigma'}]W$ vanish. Suppose
$\Delta_\sigma W \supset \int\sigma[aE_d+b\mathcal{E}]$: a change of basis by
$\mathcal{E}\to\mathcal{E}+\xi E_d$ results in shifting $a\to a+\xi b$ in the consistency conditions
that arise from terms involving $H^{ij}$. Similarly, a change of basis of $d-2$ derivative 2-index
symmetric tensors $\mathcal{H}^{ij}\to\mathcal{H}^{ij}+\xi H^{ij}$ shifts by a common amount all the
consistency conditions that arise from terms involving $H^{ij}$. So while we have not retained all
the anomalies that can contribute to the consistency conditions that lead to a potential
$C$-theorem, they give a common contribution to all those consistency conditions and therefore
effectively shift the contribution of $a$ to the potential $C$-theorem ---and the shift is
immaterial since it is basis dependent. Consider, for example, the coefficient $b$ of the anomaly
term $H^{ij}R_{ij}$ which we have retained precisely to demonstrate these points. Since
$\delta_\sigma R_{ij}= (d-2)\nabla_i\partial_j\sigma + h_{ij}\nabla^2\sigma$ it is natural to define
$\mathcal{H}_{ij}$ by
$\delta_\sigma(\sqrt{h}H^{ij}R_{ij})=\sqrt{h}[(d-2)H^{ij}+\mathcal{H}^{ij}]\nabla_i\partial_j\sigma$. With
this definition of a basis of operators the consistency conditions in Eqs.~\eqref{gen-e} below all
contain the combination $a+(d-2)b$; had we defined instead a basis with the operator
$H^{ij}R_{ij}-(d-2)E_d$ or defined the basis of 2-index tensors through
$\delta_\sigma(\sqrt{h}H^{ij}R_{ij})=\sqrt{h}\mathcal{H}^{ij}\nabla_i\partial_j\sigma$, the anomaly
$b$ would not have appeared in Eqs.~\eqref{gen-e} at all. Similarly the ambiguity due to finite
counter-terms in anomalies associated with the Lovelock tensor all enter in the combination
$A+(d-2)B$.

Imposing $[\Delta_{\sigma'},\Delta_\sigma]W=0$ we find three conditions,
\begin{subequations}
\label{gen-e}
\begin{align}
\label{gen-e1}
(\sigma\partial_{j}\sigma^{\prime}-\sigma'\partial_{j}\sigma)H^{ij}\partial_{i}X&:&& \beta^{\alpha}\partial_{\alpha}n =
   zc+a_{5\alpha}\beta^{\alpha}{ + a + (d-2)b}\\
\label{gen-e2}
(\sigma\partial_{j}\sigma^{\prime}-\sigma'\partial_{j}\sigma)H^{ij}\partial_{i}NX &:&&
     \beta^{\alpha}\partial_{\alpha}h_{1}= a +(d-2)b+2z\chi_4+\beta^{\alpha}\chi_{\alpha} -cz\\(\sigma\partial_{j}\sigma^{\prime}-\sigma'\partial_{j}\sigma)H^{ij}\partial_{i}g^{\alpha}X &:&& \partial_{\alpha}\left(a+(d-2)b\right) -\beta^{\gamma}\partial_{\gamma}a_{7\alpha}-a_{7\gamma}\partial_{\alpha}\beta^{\gamma}=za_{5\alpha}-z\chi_{\alpha}-2y_{5\alpha\gamma}\beta^{\gamma}
\end{align}
\end{subequations}
Here we have listed on the left the independent operators in  $[\Delta_{\sigma'},\Delta_\sigma]W$
whose coefficients must vanish yielding the condition correspondingly listed on  the right. We have
checked that the conditions in Eqs.~\eqref{gen-e} are invariant under the ambiguities listed in Eqs.~\eqref{gen-ambig}.
The freedom represented by these ambiguities allows us to set $a +(d-2)b+zc=0 $ in
Eq.~\eqref{gen-e1}. To see this note that $a +(d-2)b+zc$ is a vanishing anomaly per
Eq.~\eqref{gen-e1},   and Eqs.~\eqref{gen-ambig1},~\eqref{gen-ambig5} and~\eqref{gen-ambig6} give
$\delta(a +(d-2)b+zc)=-\beta^\alpha\partial_\alpha(A +(d-2)B+zC)$ which can be integrated. 
A similar argument using Eq.~\eqref{gen-e2} shows that  $a +(d-2)b+2z\chi_4-cz$ is a vanishing anomaly. 
Using this freedom we have a simpler version of the consistency conditions: 
\begin{equation*}
\begin{aligned}
(\sigma\partial_{j}\sigma^{\prime}-\sigma'\partial_{j}\sigma)H^{ij}\partial_{i}X&:&& \beta^{\alpha}\partial_{\alpha}n =a_{5\alpha}\beta^{\alpha}\\
(\sigma\partial_{j}\sigma^{\prime}-\sigma'\partial_{j}\sigma)H^{ij}\partial_{i}NX     &:&&  
 \beta^{\alpha}\partial_{\alpha}h_{1}=  \beta^{\alpha}\chi_{\alpha}
  \\                                     
(\sigma\partial_{j}\sigma^{\prime}-\sigma'\partial_{j}\sigma)H^{ij}\partial_{i}g^{\alpha}X &:&&
 \partial_{\alpha}\left(a+(d-2)b\right) -\beta^{\gamma}\partial_{\gamma}a_{7\alpha}-a_{7\gamma}\partial_{\alpha}\beta^{\gamma}=za_{5\alpha}-z\chi_{\alpha}-2y_{5\alpha\gamma}\beta^{\gamma}
\end{aligned}
\end{equation*}

Combining these we arrive at the candidate for a $C$-theorem:
\begin{equation}
\label{gen-Cthm}
\beta^{\alpha}\partial_{\alpha}\left[a+(d-2)b+zh_1-zn-\beta^{\gamma}a_{7\gamma}\right]=-2y_{5\alpha\gamma}\beta^{\gamma}\beta^{\alpha}
\end{equation}

Establishing a $C$-theorem requires in addition demonstrating
positivity of the ``metric'' $-2y_{5\alpha\gamma}$ in
Eq.~\eqref{gen-Cthm}. While we have not attempted this, it may be
possible to demonstrate this in generality working on a background
with positive definite Lovelock tensor and using the fact that
$y_{5\alpha\gamma}$ gives the RG response of the contact counter-term
to the obviously positive definite correlator
$\langle \mathcal {O}_\alpha\mathcal{O}_\gamma\rangle$. In addition,
one should check that, when computed at the gaussiaan fixed point, the
quantity $a+(d-2)b+zh_1-zn-\beta^{\gamma}a_{7\gamma}$ is a measure of
the number of degrees of freedom. We hope to come back to this
questions in the future, by performing explicit calculations (at and
away from fixed points) of these quantities --- but such extensive
computations are beyond the scope of this work.

The limit $d=2$ is special since $H^{ij}=2h^{ij}$. In our analysis,
the term $H^{ij}R_{ij}=2R=2E_2$ so $a$ and $b$ appear in the
combination $a+2b$ throughout. The potential $C$ theorem reads
\begin{equation}
\label{gen-2dim}
\beta^{\alpha}\partial_{\alpha}\left[a+2b+zh_1-zn-\beta^{\gamma}a_{7\gamma}\right]=-2y_{5\alpha\gamma}\beta^{\gamma}\beta^{\alpha}
\end{equation}
As we have seen in Sec.~\ref{sec:allz}, potential $C$-theorems in
$d=2$ for any $z$ can be found only in the $\nabla^2\partial_t$
sector. Consulting Tab.~\ref{tab:TXX} we see the only candidate for
$X$ in our present discussion is $X=K$. None of the potential
$C$-theorems listed in Eqs.~\eqref{CcandXXT1}--\eqref{CcandXXT4} (nor
linear combinations thereof) reproduce the potential $C$-theorem in
Eq.~\eqref{gen-2dim}. The reason for this is that in
Sec.~\ref{subsec:apps} we looked for $C$-theorems from consistency
conditions that included, among others, tems with
$\sigma\nabla_i\partial_j\sigma'-\sigma'\nabla_i\partial_j\sigma$,
wheras in this section we integrated such terms by parts. The
difference then corresponds to 
combining the consistency conditions  given in the  appendix with some
of their derivatives.

In fact we have found a scheme for deducing aditional $C$-theorem
candidates in $d=2$ by taking derivatives of some of our consistency
conditions. The method is as follows. Take $X\in
\{R,\nabla^{2}N,\partial_{i}N\partial^{i}N,K\}$; the first three
instances apply to the case $z=2$ while the last is applicable for
arbitrary $z$. Then : 
\begin{itemize}
\item Consider  the consistency condition involving
  $\sigma\nabla^{2}\sigma^{\prime}X$, and take a derivative to obtain
  an equation, say $T_1$.
\item Take the consistency condition involving
  $\sigma\nabla_{i}\sigma^{\prime}\partial^{i}NX$. From this one may deduce a linear combination of anomalies is vanishing.  Set
  that to $0$ using the ambiguity afforded by counter-terms. The
  remaining terms in the equation (all proportional to $\beta^\alpha$)
  give an equation we denote by $T_2$.
\item Take the consistency condition involving
  $\sigma\nabla_{i}\sigma^{\prime}\partial^{i}g^{\alpha}X$, contract
  it with $\beta^{\alpha}$, to get an equation, say,  $T_{3}$.
\item Combine $T_1,T_2,T_3$ in a manner such that there are no terms
  of the form $\beta^{\alpha}r_{\alpha\cdots}$ and $r_{\gamma\cdots}\beta^{\alpha}\partial_{\alpha}\beta^{\gamma}$.
\end{itemize}
Following this scheme we obtain four new $C$-theorem candidates. In
the following the expressions for $T_{1,2,3}$ refer to the equation
numbers of the consistency conditions in the appendix:
\begin{enumerate}[(i)]
\item $X=R$. $T_1= \text{\ref{eq:del2R}}, T_2={\ref{eq:dsRdN}}, T_3= \text{\ref{eq:dsRdg}}$. Set
  $c-\chi_{4}=0$. Then 
\begin{equation}
\label{c-der-1}
\beta^{\alpha}\partial_{\alpha}\left[8a+2c+2h_1+2\beta^{\gamma}\partial_{\gamma}n-\beta^{\gamma}a_{7\gamma}\right]=2\beta^{\alpha}\beta^{\gamma}\left[\partial_{\alpha}a_{5\gamma}-y_{5\alpha\gamma}\right]
\end{equation}
\item $X=\nabla^{2}N$.  $T_1=\text{\ref{eq:del2N}}, 
  T_2=\text{\ref{eq:dsdNd2N}}, T_3=\text{\ref{eq:dsdgdel2N}}$. Set $4p_{4}+8\rho_{23}=0$. Then 
\begin{equation}
\label{c-der-2}
\beta^{\alpha}\partial_{\alpha}\left[8\rho_{23}+4c+2\rho_{12}+2\beta^{\gamma}\partial_{\gamma}h_2-\beta^{\gamma}\rho_{13\gamma}\right]=2\beta^{\alpha}\beta^{\gamma}\left[\partial_{\alpha}\rho_{24\gamma}-x_{2\alpha\gamma}\right]
\end{equation}
\item $X=\nabla_{i}N\nabla^{i}N$. $T_1=\text{\ref{eq:dsdNdN}},
  T_2=\text{\ref{eq:dsdNdNdN}} , T_3=\text{\ref{eq:dsdNdNdg}}$. 
Set  $8p_{3}+4p_{4}=0$. Then 
\begin{equation}
\label{c-der-3}
\beta^{\alpha}\partial_{\alpha}\left[4\chi_{4}-4p_{4}+2\rho_{11}+2\beta^{\gamma}\partial_{\gamma}\chi_3-\beta^{\gamma}\rho_{8\gamma}\right]=2\beta^{\alpha}\beta^{\gamma}\left[\partial_{\alpha}y_{\gamma}-x_{\alpha\gamma}\right]
\end{equation}
\item $X=K$.  $T_1=\text{\ref{w2}}, T_2=\text{\ref{u1}}, T_3=\text{\ref{eq:dsKdg}}$.
Set  $j-\rho_{4}=0$. Then 
\begin{equation}
\label{same-as-gral}
\beta^{\alpha}\partial_{\alpha}\left[4b+zj+zl_{1}+2\beta^{\gamma}\partial_{\gamma}m-\beta^{\gamma}\rho_{7\gamma}\right]=2\beta^{\alpha}\beta^{\gamma}\left[\partial_{\alpha}b_{8\gamma}-x_{5\alpha\gamma}\right]
\end{equation}
\end{enumerate}
We have verified that after accounting for differences in basis and
notation Eq.~\eqref{same-as-gral} is precisely the same as the
general $C$-theorem candidate of this section given  in Eq.~\eqref{gen-2dim}.

\section{Summary and Discussion}
\label{sec:conc}
Wess-Zumino consistency conditions for Weyl transformations impose constraints on the
renormalization group flow of Weyl anomalies. As a first step in studying these
constraints in non-relativistic quantum field theories we have classified the anomalies
that appear in $d=2$ (spatial dimensions) at $z=2$ (dynamical exponent at gaussian fixed
point). There are many more anomalies than in the comparable relativistic case (3+1
dimensions): there are 39
anomalies associated with 4-spatial derivatives (Table~\ref{tab:XXXX}), 6 with 2-time
derivatives (Table~\ref{tab:TT}) and 32 more that contain 1-time and 2-spatial derivaties
(Table~\ref{tab:TXX}). Freedom to add finite amounts to  counterterms gives in turn
freedom to shift some anomalies arbitrarily. ``Trivial Anomalies'' are those that  can
thus be set to zero. We  then classified all counterterms
(Tables~\ref{tab:XXXX}--\ref{tab:TXX}), gave  the shift in Weyl
anomalies produced by shifts in counterterms (in App.~\ref{app:anom-ambig}),  
and then listed the trivial anomalies (Table~\ref{tab:triv}). 

The consistency conditions among these $39+6+32$ anomalies do not mix among the three
sectors. They are listed by sector in App.~\ref{app:conditions22}, and from these we can
read-off ``Vanishing Anomalies'' --- those that vanish at fixed points; see
Table.~\ref{tab:vanishing}. As an application of the use of these conditions we find 6
combinations that give $C$-function candidates. That is, we find (combinations of)
anomalies $\tilde a$ and $\mathcal{H}_{\alpha\beta}$ that satisfy
\(\mu\, d\tilde{a}/d\mu=\mathcal{H}_{\alpha\beta}\beta^{\alpha}\beta^{\beta}\),
where \( \beta^\alpha= \mu\, dg^\alpha/d\mu\)
give the flow of the dimensionless coupling constants; then $\tilde a$ flows monotonically
provided $\mathcal{H}_{\alpha\beta}$ is positive definite. We have not endeavored to
attempt to prove that any of our $\mathcal{H}_{\alpha\beta}$ functions are positive
definite, and hence our candidates remain just that, candidates. Exploring positivity of
these functions in specific examples would be of interest, and determining
model-independently under which conditions positivity holds would be more so.

It is important to appreciate the generality, or lack of it thereof, of our results. While
we have used some specific form of the Lagrangian in setting up and contextualizing the
computation, there is in fact no need to assume this in order to classify the anomalies
and compute the consistency conditions. On the other hand we have made a fairly strong
assumption, that the classical action integral is invariant under the anisotropic scale
transformation $\vec x \mapsto \lambda \vec x, t \mapsto \lambda^{z} t$. All our couplings
correspond to marginal deformations. In the $3+1$-dimensional relativistic case relevant
deformations do modify the consistency conditions, but the candiate $C$-theorem is not
affected, at least by a class of relevant deformations~\cite{Osborn:1991gm}. Clearly,
another interesting direction of future study is to investigate the effect of relevant
deformations on our consistency conditions: perhaps some of the~6 $C$-candidates survive
even in the presence of relevant deformations, much as in the relativistic case.

While we have performed a detailed analysis only for the $z=2$ case in $2+1$ dimensions,
our results can be readily used in other cases too. For theories in $2+1$ dimensions with
$z>0$ and neither $z=2k$ nor $z=2/k$ where $k$ is an integer, only the sector of anomalies with
1-time and 2-spatial derivatives remains. Moreover, the classification of anomalies and
the consistency conditions for that sector that were derived assuming $z=2$ are valid for
arbitrary $z$, with minor modifications in the form of a sprinkling of factors of $z/2$;
we have retained explicit $z$ dependence in the consistency
conditions in this sector, Eqs.~\eqref{eqs:TXXconditions}. This means, in
particular, that the~4 $C$-candidates in this sector, in
Eqs.~\eqref{CcandXXT1}--\eqref{CcandXXT4}, are $C$-candidates for arbitrary $z$. For 
$z=2k\ge4$ there are anomalies with $2(k+1)$ spatial derivatives; their classification depends on $z$,
so a case-by-case analysis is required. For $z=2/k\le2$ there are anomalies with $k+1$
time derivatives; again  their classification depends on $z$ and a case-by-case analysis
is required. 

For spatial dimensions $d>2$, if $d$ is even a $C$-theorem candidate, in Eq.~\eqref{gen-2dim},
becomes available that mimics the one in relativistic theories. Again it relies on assuming only
marginal operators are present, but it is possible that, just as in the $3+1$ realtivistic case, the
conclusion is not modified by inclusion of relevant deformations.  The candidate is based on the
anomaly associated with the $d$-dimensional Euler density for the theory on a curved
background. Here again it would be interesting to have an explicit example, to test whether the
putative metric in coupling constant space, $\mathcal{H}_{\alpha\beta}$, is positive definite.  The
analysis of a potential $C$-theorem in the case of general dimensions $d$ yields four additional
potential $C$-theorems in $d=2$, three for $z=2$ given in Eqs.~\eqref{c-der-1}--\eqref{c-der-3} and
one more for arbitrary $z$, given in Eq.~\eqref{same-as-gral}.  It deserves mention that all of our
proposed $C$ theorem candidates are scheme dependent even at a fixed point. Hence, the value of them
at a fixed point can be shifted using counter-terms $F$.

If any of these candidates yields a bona-fide $C$-theorem the presence
of limit cycles in non-relativistic quantum field theories is called
into question. Limit cycles in relativistic 3+1 dimensional theories
physically correspond to critical points, and the recursive flow
corresponds to what amounts to a simultaneous rotation among
fundamental fields and marginal operators and their
coefficients. Cyclic behavior in non-relativistic quantum systems, on
the other hand, do not display continuous scale invariance, so there
is no reason to expect that $C$ would remain constant along the
flow. The resolution may be that there are no $C$-theorems at all. Or
that there are $C$-theorems only under conditions that do not apply to
systems that exhibit cycles. We look forward to developments in this
area.

\begin{acknowledgments}
  We would like to thank John McGreevy for insightful comments. SP would like to thank Andreas Stergiou for help with Mathematica.  This work was supported in part by the US Department of Energy under contract DE- SC0009919.
\end{acknowledgments}

\appendix

\section{Consistency Conditions for $2+1$d NRCFT}
\label{app:conditions22}
We give below the consistency conditions for the $d=2$ theory. In the $\partial_t\nabla^2$ sector they are given  for
arbitrary $z$; else $z=2$ is assumed.   The conditions in the  $\partial_t\nabla^2$ sector are as follows: 
\begin{subequations}
\label{eqs:TXXconditions}
\begin{align}
\label{u0}
   \sigma\partial_{t}\sigma^{\prime} \partial_{i}N\partial^{i}g^{\alpha}&:
 &&-\beta^{\gamma}\partial_{\gamma}\rho_{1\alpha}-\rho_{1\gamma}\partial_{\alpha}\beta^{\gamma}+2\rho_{\alpha}-2\partial_{\alpha}j+p_{4\gamma\alpha}\beta^{\gamma}=0
 \\
 \label{w0}
   \sigma\partial_{t}\sigma^{\prime}\nabla^{2}g^{\alpha}&:
&&-b_{4\sigma}\partial_{\alpha}\beta^{\sigma}-\beta^{\sigma}\partial_{\sigma}b_{4\alpha}+x_{4\gamma\alpha}\beta^{\gamma}+2b_{8\alpha}=0
\\
\nabla^{2}\sigma\partial_{t}\sigma^{\prime} &:
&&2k +2m -zl_{2}-b_{4\alpha}\beta^{\alpha}+\beta^{\alpha}b_{9\alpha}=0
\\
\label{w1}
-\sigma^{\prime}\partial_{t}\sigma\nabla^{2}N&:
&& 2j+\beta^{\alpha}\partial_{\alpha}l_2=b_{6\alpha}\beta^{\alpha}
\\
\label{q2}
\sigma^{\prime}\partial_{i}\sigma\partial^{i}N\partial_{t}g^{\alpha}&:
&&-2zb_{6\alpha}-2z\rho_{6\alpha}+\rho_{5\gamma}\partial_{\alpha}\beta^{\gamma}+\beta^{\gamma}\partial_{\gamma}\rho_{5\alpha}-\beta^{\gamma}p_{4\alpha\gamma}=0
\\
\label{w2}
\sigma^{\prime}\nabla^{2}\sigma K &:
&&2b+\beta^{\alpha}\partial_{\alpha}m +zj=\beta^{\alpha}b_{8\alpha}
\\
\sigma^{\prime}\partial_{t}\sigma R &:
&&2b-\beta^{\alpha}\partial_{\alpha}k+b_{5\alpha}\beta^{\alpha}=0
\\
\label{eq:dsKdg}
\sigma^{\prime}\partial_{i}\sigma   K\partial^{i}g^{\alpha}&:
&&-2x_{5\alpha\gamma}\beta^{\gamma}+\beta^{\gamma}\partial_{\gamma}b_{7\alpha}-z\rho_{\alpha}\nonumber\\
& &&\qquad+b_{7\gamma}\partial_{\alpha}\beta^{\gamma}-2b_{8\gamma}\partial_{\alpha}\beta^{\gamma}+z\partial_{\alpha}j=0\
\\
\sigma^{\prime}\nabla^{2}\sigma\partial_{t}g^{\alpha}&:
&&-x_{4\alpha\gamma}\beta^{\gamma}+2b_{5\alpha}-zb_{6\alpha}+b_{9\gamma}\partial_{\alpha}\beta^{\gamma}+\beta^{\gamma}\partial_{\gamma}b_{9\alpha}=0\
\\
\partial_{i}\sigma\partial_{t}\sigma^{\prime}\partial^{i}g^{\alpha}&:
&&-2b_{4\gamma}\partial_{\alpha}\beta^{\gamma}+2b_{7\alpha}+x_{6\gamma\alpha}\beta^{\gamma}-2b_{3\alpha\gamma}\beta^{\gamma}-z\rho_{1\alpha}=0 
\\
\sigma\partial_{t}\sigma^{\prime}\partial_{i}g^{\alpha}\partial^{i}g^{\beta}&:
&&-\beta^{\gamma}\partial_{\gamma}b_{3\alpha\beta}-b_{3\gamma\beta}\partial_{\alpha}\beta^{\gamma}\nonumber\\
& &&\qquad-b_{3\gamma\alpha}\partial_{\beta}\beta^{\gamma}-b_{4\gamma}\partial_{\alpha}\partial_{\beta}\beta^{\gamma}+x_{3\gamma\alpha\beta}\beta^{\gamma}+2x_{5\alpha\beta}=0
\\
-\sigma\partial_{i}\sigma^{\prime}\partial_{t}g^{\alpha}\partial^{i}g^{\beta}&:
&&-2x_{3\alpha\gamma\beta}\beta^{\gamma}+x_{6\sigma\beta}\partial_{\alpha}\beta^{\sigma}\nonumber\\
& &&\qquad+x_{6\alpha\gamma}\partial_{\beta}\beta^{\gamma}-2x_{4\alpha\sigma}\partial_{\beta}\beta^{\sigma}+\beta^{\gamma}\partial_{\gamma}x_{6\alpha\beta}-zp_{4\alpha\beta}=0
\\
\label{q3}
\sigma\partial_{t}\sigma^{\prime}\frac{\partial^{i}N}{N}\frac{\partial_{i}N}{N}&:
&&\beta^{\gamma}\partial_{\gamma}\rho_{3}-2\rho_{4}-\beta^{\alpha}\rho_{6\alpha}=0
\\
\label{u1}
\sigma\partial_{i}\sigma^{\prime}K\partial^{i}N&:
&&2zj-\beta^{\alpha}\rho_\alpha+\beta^{\alpha}\partial_{\alpha}l_{1}-2z\rho_{4}=0
\\
\label{so0}
\partial_{i}\sigma\partial_{t}\sigma^{\prime}\partial^{i}N&:
&&2z\rho_{3}-2l_1+2zl_2+2j+\rho_{1\alpha}\beta^{\alpha}=0
\\
\label{so1}
\sigma^{\prime}\partial^{j}\sigma\partial^{i}N\left(K_{ij}-\tfrac{1}{2}Kh_{ij}\right)&:
&&2zf_1-\beta^{\alpha}f_{3\alpha}+zf_4-\beta^{\alpha}\partial_{\alpha}f_{7}=0\\
\label{so2}
\sigma^{\prime}\partial^{j}\sigma\partial^{i}g^{\alpha}\left(K_{ij}-\tfrac{1}{2}Kh_{ij}\right)&:
&&2f_{2\alpha\gamma}\beta^{\gamma}+zf_{3\alpha}-\beta^{\gamma}\partial_{\gamma}f_{8\alpha}-f_{8\gamma}\partial_{\alpha}\beta^{\gamma} +2f_{5\gamma}\partial_{\alpha}\beta^{\gamma}=0\\
\label{so3}
\sigma^{\prime}\nabla^{i}\partial^{j}\sigma\left(K_{ij}-\tfrac{1}{2}Kh_{ij}\right)&:
&&zf_4+\beta^{\alpha}f_{5\alpha}-\beta^{\alpha}\partial_{\alpha}f_{6}=0
\end{align}
\end{subequations}

The conditions coming from $\partial_t^2$ sector are as follows:
\begin{subequations}
\label{eqs:TTconditions}
\begin{align}
\label{dtK}
\sigma^{\prime}\partial_{t}\sigma K &:
&&4d- \beta^{\alpha}\partial_{\alpha}f +\beta^{\alpha}w_\alpha
  =0 
\\
\label{dt2g}
\sigma^{\prime}\partial_{t}\sigma\partial_{t}g^{\alpha}&:
&&-2w_{\alpha}+\beta^{\gamma}\partial_{\gamma}b_{\alpha}+b_{\gamma}\partial_{\alpha}\beta^{\gamma}-2\chi_{0\alpha\gamma}\beta^{\gamma}=0
\end{align}
\end{subequations}

The conditions coming from the $\nabla^4$ sector are given by:
\begin{subequations}
\label{eqs:XXXXconditions}
\begin{align}
\partial_{i}\sigma^{\prime}\nabla^{2}\sigma \partial^{i}g^{\alpha}&:
&&-2\rho_{13\alpha}-\beta^{\gamma}\rho_{21\alpha\gamma}+2a_{7\alpha}\nonumber\\
& &&\qquad+2\chi_{1\alpha}+2a_{3\alpha\gamma}\beta^{\gamma}+2a_{4\gamma}\partial_{\alpha}\beta^{\gamma}=0
\\
\sigma^{\prime}\nabla^{2}\sigma\partial_{i} g^{\alpha} \partial^{i} g^{\beta} &:
&&-\beta^{\gamma}t_{2\gamma\alpha\beta}+\beta^{\gamma}\partial_{\gamma}a_{3\alpha\beta}+a_{3\alpha\gamma}\partial_{\beta}\beta^{\gamma}\nonumber\\
& &&\qquad+a_{3\beta\gamma}\partial_{\alpha}\beta^{\gamma}+a_{4\gamma}\partial_{\alpha}\partial_{\beta}\beta^{\gamma}+2y_{5\alpha\beta}-2x_{2\alpha\beta}=0
\\
\label{eq:del2sigdel2g}
\sigma^{\prime}\nabla^{2}\sigma
  \nabla^{2}g^{\alpha}&:
&&2a_{5\alpha}+\beta^{\gamma}\partial_{\gamma}a_{4\alpha}+a_{4\gamma}\partial_{\alpha}\beta^{\gamma}-2\rho_{22\alpha\gamma}\beta^{\gamma}-2\rho_{24\alpha}=0
\\ 
\label{eq:dsdgdel2N}
\sigma^{\prime}\partial_{i}\sigma\partial^{i}g^{\alpha}\nabla^{2}N&:
&&-2x_{2\alpha\gamma}\beta^{\gamma}+\beta^{\gamma}\partial_{\gamma}\rho_{13\alpha}+\rho_{13\gamma}\partial_{\alpha}\beta^{\gamma}\nonumber\\
& &&\qquad-2\rho_{24\gamma}\partial_{\alpha}\beta^{\gamma}-2\rho_{25\alpha}=0
\\
\sigma^{\prime}\partial_{i}\sigma\partial^{i}N \partial_{j}N\partial^{j}g^{\alpha}&:
&&-2p_{5\beta\alpha}\beta^{\beta}+\beta^{\gamma}\partial_{\gamma}\rho_{10\alpha}+\rho_{10\gamma}\partial_{\alpha}\beta^{\gamma}-4\rho_{25\alpha}-4\rho_{9\alpha}=0
\\
\sigma^{\prime}\partial_{i}\sigma\partial^{i}g^{\alpha}\partial_{j}g^{\beta}\partial^{j}g^{\gamma}&:
&&-4x_{\alpha\sigma\beta\gamma}\beta^{\sigma}+\beta^{\sigma}\partial_{\sigma}t_{\alpha\beta\gamma}+t_{\sigma\beta\gamma}\partial_{\alpha}\beta^{\sigma}+t_{\alpha\sigma\gamma}\partial_{\beta}\beta^{\sigma}
+t_{\alpha\sigma\beta}\partial_{\gamma}\beta^{\sigma}\nonumber\\
& &&\qquad-2t_{2\sigma\beta\gamma}\partial_{\alpha}\beta^{\sigma}
+\rho_{21\alpha\sigma}\partial_{\beta}\partial_{\gamma}\beta^{\sigma}-2x_{\alpha\beta\gamma}=0
\\
\sigma^{\prime}\partial_{i}\sigma\partial^{i}g^{\alpha}\partial_{j}N\partial^{j}g^{\beta}
&:
&&-4p_{5\alpha\beta}-2\rho_{26\beta\gamma}\partial_{\alpha}\beta^{\gamma}
\nonumber\\
& &&\qquad+x_{1\alpha\gamma}\partial_{\beta}\beta^{\gamma}+x_{1\gamma\beta}\partial_{\alpha}\beta^{\gamma}+\beta^{\gamma}\partial_{\gamma}x_{1\alpha\beta}-2x_{\alpha\gamma\beta}\beta^{\gamma}=0
\\
\sigma\partial_{i}\sigma^{\prime}\partial^{i}g^{\alpha}\nabla^{2}g^{\beta}&:
&&-\beta^{\gamma}\partial_{\gamma}\rho_{21\alpha\beta}-\rho_{21\gamma\beta}\partial_{\alpha}\beta^{\gamma}-\rho_{21\alpha\gamma}\partial_{\beta}\beta^{\gamma}\nonumber\\
& &&\qquad+4\rho_{22\gamma\beta}\partial_{\alpha}\beta^{\gamma}+2\rho_{26\alpha\beta}+2t_{2\beta\gamma\alpha}\beta^{\gamma}=0\
\\
\label{eq:del2N}
\sigma^{\prime}\nabla^{2}\sigma\nabla^{2}N&:
&&-4\rho_{23}-\beta^{\gamma}\rho_{24\gamma}+\beta^{\gamma}\partial_{\gamma}h_{2}-2c=0\
\\
\label{eq:dsdNd2N}
\sigma^{\prime}\partial_{i}\sigma \partial^{i}N\nabla^{2}N&:
&&4p_4-\beta^{\alpha}\partial_{\alpha}\rho_{12}+8\rho_{23}+\beta^{\gamma}\rho_{25\gamma}=0
\\
\label{eq:dsdNd2g}
\sigma^{\prime}\partial_{i}\sigma\partial^{i}N\nabla^{2}g^{\alpha}&:
&&-4y_{\alpha}+\beta^{\gamma}\partial_{\gamma}\rho_{7\alpha}+\rho_{7\gamma}\partial_{\alpha}\beta^{\gamma}-4\rho_{24\alpha}-\beta^{\gamma}\rho_{26\gamma\alpha}=0
\\
\label{eq:del2R}
-\sigma^{\prime}\nabla^{2}\sigma R&:
&&-a_{5\alpha}\beta^{\alpha}+4a+2c+\beta^{\alpha}\partial_{\alpha}n=0
\\
\label{eq:dsRdg}
\sigma^{\prime}\partial_{i}\sigma
  R\partial^{i}g^{\alpha}&:
&&-2y_{5\alpha\gamma}\beta^{\gamma}-2\chi_{\alpha}+2\partial_{\alpha}c \nonumber\\
& &&\qquad+\beta^{\gamma}\partial_{\gamma}a_{7\alpha}+a_{7\gamma}\partial_{\alpha}\beta^{\gamma}-2a_{5\gamma}\partial_{\alpha}\beta^{\gamma}=0
\\
\label{eq:d2sdgdN}
\sigma\nabla^{2}\sigma^{\prime}\partial^{i}g^{\alpha}\partial_{i}N&:
&&-\beta^{\gamma}\partial_{\gamma}\chi_{1\alpha}-\chi_{1\gamma}\partial_{\alpha}\beta^{\gamma}+2\partial_{\alpha}c-2\chi_{\alpha}+\rho_{26\alpha\gamma}\beta^{\gamma}+2\rho_{25\alpha}=0
\\
\label{eq:dsd2sdN}\partial_{i}\sigma\nabla^{2}\sigma^{\prime}\partial^{i}N&:
&& 2h_1+4h_2-2c+\beta^{\alpha}\chi_{1\alpha}+4\chi_{3}-\beta^{\alpha}\rho_{7\alpha}-2\rho_{12}=0
\\
\label{eq:dsdNdNdN}
\sigma^{\prime}\partial_{i}\sigma \partial^{i}N\partial_{j}N\partial^{j}N&:
&&
8p_{3}-\beta^{\alpha}\partial_{\alpha}\rho_{11}+\rho_{9\alpha}\beta^{\alpha}+4p_{4}=0
\\
\sigma\partial_{i}\sigma^{\prime}\partial_{j}g^{\alpha}\partial^{j}g^{\beta}\partial^{i}N&:
&&
4x_{\alpha\beta}-\rho_{7\gamma}\partial_{\beta}\partial_{\alpha}\beta^{\gamma}-\rho_{1\gamma\beta}\partial_{\alpha}\beta^{\gamma}\nonumber\\
& &&\qquad-\rho_{1\gamma\alpha}\partial_{\beta}\beta^{\gamma}-\beta^{\gamma}\partial_{\gamma}\rho_{1\alpha\beta}+\beta^{\gamma}x_{\gamma\alpha\beta}+4x_{2\alpha\beta}=0
\\
\label{eq:dsRdN}
\sigma\partial_{i}\sigma^{\prime}R\partial^{i}N&:
&&4c+\beta^{\alpha}\partial_{\alpha}h_1-\beta^{\alpha}\chi_{\alpha}-4\chi_{4}=0
\\
\label{eq:dsdNdN}
  \sigma^{\prime}\nabla^{2}\sigma\partial^{i}N\partial_{i}N&:
&&2\chi_{4}+\beta^{\alpha}\partial_{\alpha}\chi_{3}-\beta^{\alpha}y_{\alpha}-2p_{4}=0
\\
\label{eq:dsdNdNdg}
\sigma^{\prime}\partial_{i}\sigma\partial_{j}N\partial^{j}N\partial^{i}g^{\alpha}&:
&&-2x_{\alpha\gamma}\beta^{\gamma}+\beta^{\gamma}\partial_{\gamma}\rho_{8\alpha}+\rho_{8\gamma}\partial_{\alpha}\beta^{\gamma}-2\rho_{9\alpha}-2y_{\gamma}\partial_{\alpha}\beta^{\gamma}=0
\end{align}
\end{subequations}

\section{Anomaly ambiguities}
\label{app:anom-ambig}
As explained in Sec.~\ref{sec:CTs} the freedom to shift counter-terms by finite amount makes anomaly
coefficients ambiguous.  We list here the precise form of these ambiguities, in the $\partial_t\nabla^2$ sector they are given  for
arbitrary $z$; else $z=2$ is assumed :

\subsection{$\partial_{t}^{2}$ Sector}
\begin{subequations}
\begin{align}
\label{eq:f-ambig}
\delta f &=-4D-\beta^{\alpha}W_{\alpha}\\
\delta w_{\alpha}&= -\left[\beta^{\gamma}\partial_{\gamma}W_{\alpha}+W_{\gamma}\partial_{\alpha}\beta^{\gamma}\right]\\
\label{eq:balpha-ambig}
\delta b_{\alpha}&=-2W_{\alpha}-2X_{0\alpha\gamma}\beta^{\gamma}\\
\delta \chi_{0\alpha\beta}&=-\beta^{\gamma}\partial_{\gamma}X_{0\alpha\beta}-X_{0\alpha\gamma}\partial_{\beta}\beta^{\gamma}-X_{0\alpha\gamma}\partial_{\beta}\beta^{\gamma}\\
\label{eq:d-ambig}
\delta d &= -\beta^{\alpha}\partial_{\alpha}D\\
\delta e &= -\beta^{\alpha}\partial_{\alpha}E
\end{align}
\end{subequations}

\subsection{$\partial_{t}\nabla^{2}$ Sector}
\begin{subequations}
\begin{align}
\delta\rho_{4}&= -\beta^{\alpha}\partial_{\alpha}P\\
\delta x_{5\alpha\beta}&=-P_{3\gamma}\partial_{\alpha}\partial_{\beta}\beta^{\gamma}-\beta^{\gamma}\partial_{\gamma}X_{5\alpha\beta}-X_{5\gamma\beta}\partial_{\alpha}\beta^{\gamma}-X_{5\gamma\alpha}\partial_{\beta}\beta^{\gamma}\\
\delta\rho_{\alpha}&=-\beta^{\gamma}\partial_{\gamma}P_{\alpha}-P_{\gamma}\partial_{\alpha}\beta^{\gamma}\\
\delta j &= -\beta^{\alpha}\partial_{\alpha}L\\
\delta b_{8\alpha}&=-\beta^{\gamma}\partial_{\gamma}P_{3\alpha}-P_{3\gamma}\partial_{\alpha}\beta^{\gamma}\\
\delta b&= -\beta^{\alpha}\partial_{\alpha}B\\
\delta m&=2B+zL-P_{3\alpha}\beta^{\alpha}\\
\delta l_{1}&= -2zP+2zL-\beta^{\alpha}P_{\alpha}\\
\delta b_{7\alpha}&=-2P_{3\gamma}\partial_{\alpha}\beta^{\gamma}+z\partial_{\alpha}L-zP_{\alpha}-2X_{5\alpha\gamma}\beta^{\gamma}\\
\delta\rho_{6\alpha}&=-X_{\gamma}\partial_{\alpha}\beta^{\gamma}-\beta^{\gamma}\partial_{\gamma}X_{\alpha}\\
\delta x_{3\alpha\beta\gamma}&=-\beta^{\sigma}\partial_{\sigma}X_{3\alpha\beta\gamma}-X_{3\sigma\beta\gamma}\partial_{\alpha}\beta^{\sigma}-X_{3\alpha\sigma\gamma}\partial_{\beta}\beta^{\sigma}-X_{3\alpha\sigma\beta}\partial_{\gamma}\beta^{\sigma}-X_{4\alpha\sigma}\partial_{\gamma}\partial_{\beta}\beta^{\sigma}\\
\delta p_{4\alpha\beta}&=-\beta^{\gamma}\partial_{\gamma}P_{4\alpha\beta}-P_{4\gamma\beta}\partial_{\alpha}\beta^{\gamma}-P_{4\alpha\gamma}\partial_{\beta}\beta^{\gamma}\\
\delta b_{6\alpha}&=-\beta^{\gamma}\partial_{\gamma}B_{6\alpha}-B_{6\gamma}\partial_{\alpha}\beta^{\gamma}\\
\delta x_{4\alpha\beta}&=-X_{4\gamma\beta}\partial_{\alpha}\beta^{\gamma}-\beta^{\gamma}\partial_{\gamma}X_{4\alpha\beta}-X_{4\alpha\gamma}\partial_{\beta}\beta^{\gamma}\\
\delta b_{5\alpha}&=-\beta^{\gamma}\partial_{\gamma}B_{5\alpha}-B_{5\gamma}\partial_{\alpha}\beta^{\gamma}\\
\delta b_{9\alpha}&=2B_{5\alpha}-zB_{6\alpha}-\beta^{\gamma}X_{4\alpha\gamma}\\
\delta \rho_{5\alpha}&=-2zX_{\alpha}-2zB_{6\alpha}-P_{4\alpha\gamma}\beta^{\gamma}\\
\delta x_{6\alpha\beta}&=-X_{4\alpha\gamma}\partial_{\beta}\beta^{\gamma}-2X_{3\alpha\gamma\beta}\beta^{\gamma}-zP_{4\alpha\beta}\\
\delta\rho_{3}&=-2P-\beta^{\alpha}X_{\alpha}\\
\delta b_{3\alpha\beta}&=-2X_{5\alpha\beta}-X_{3\gamma\beta\alpha}\beta^{\gamma}\\
\delta\rho_{1\alpha}&=-2P_{\alpha}+\partial_{\alpha}2L-P_{4\gamma\alpha}\beta^{\gamma}\\
\delta l_2&=2L-B_{6\alpha}\beta^{\alpha}\\
\delta b_{4\alpha}&=-2P_{3\alpha}-X_{4\gamma\alpha}\beta^{\gamma}\\
\delta k&=-2B-B_{5\alpha}\beta^{\alpha}\\
\delta f_1 &= -\beta^{\alpha}\partial_{\alpha}F_1\\
\delta f_{2\alpha\beta}&= -\beta^{\gamma}\partial_{\gamma}F_{2\alpha\beta}- F_{2\gamma\beta}\partial_{\alpha}\beta^{\gamma}-F_{2\alpha\gamma}\partial_{\beta}\beta^{\gamma}-F_{5\gamma}\partial_{\alpha}\partial_{\beta}\beta^{\gamma}\\
\delta f_{3\alpha}&= -\beta^{\gamma}\partial_{\gamma}F_{3\alpha}-F_{3\gamma}\partial_{\alpha}\beta^{\gamma}\\
\delta f_{4}&=-\beta^{\gamma}\partial_{\gamma}F_4\\
\delta f_{5\alpha}&=-\beta^{\gamma}\partial_{\gamma}F_{5\alpha}-F_{5\gamma}\partial_{\alpha}\beta^{\gamma}\\
\delta f_{6}&=-zF_4-F_{5\alpha}\beta^{\alpha}\\
\delta f_{7}&=-2zF_1-F_{3\alpha}\beta^{\alpha}-zF_4\\
\delta f_{8\alpha}&=-zF_{3\alpha}-2F_{2\gamma\alpha}\beta^{\gamma}-2F_{5\gamma}\partial_{\alpha}\beta^{\gamma}
\end{align}
\end{subequations}

\subsection{$\nabla^{4}$ Sector}
\begin{subequations}
\begin{align}
\delta p_{3}&=-\beta^{\alpha}\partial_{\alpha}P_{3}\\
\delta x_{\alpha\beta}&=-Y_{\gamma}\partial_{\beta}\partial_{\alpha}\beta^{\gamma}-X_{\gamma\beta}\partial_{\alpha}\beta^{\gamma}-X_{\gamma\alpha}\partial_{\beta}\beta^{\gamma}-\beta^{\gamma}\partial_{\gamma}X_{\alpha\beta}\\
\delta \rho_{9\alpha}&=-\beta^{\gamma}\partial_{\gamma}P_{1\alpha}-P_{1\gamma}\partial_{\alpha}\beta^{\gamma}\\
\label{eq:p4-ambig}
\delta p_{4}&=-\beta^{\alpha}\partial_{\alpha}P_{4}\\
\delta y_{\alpha}&=-\beta^{\gamma}\partial_{\gamma}Y_{\alpha}-Y_{\gamma}\partial_{\alpha}\beta^{\gamma}\\
\label{eq:chi-ambig}
\delta\chi_{4}&= -\beta^{\alpha}\partial_{\alpha}Q\\
\delta\chi_{3}&=2Q -\beta^{\alpha}Y_{\alpha}-2P_4\\
\delta\rho_{11}&=-8P_3-P_{1\alpha}\beta^{\alpha}-4P_{4}\\
\delta\rho_{8\alpha}&=-2P_{1\alpha}-2X_{\alpha\gamma}\beta^{\gamma}-2Y_{\gamma}\partial_{\alpha}\beta^{\gamma}\\
\delta x_{\alpha\beta\gamma\delta}&=-\beta^{\sigma}\partial_{\sigma}X_{\alpha\beta\gamma\delta}-X_{\sigma\beta\gamma\delta}\partial_{\alpha}\beta^{\sigma}-X_{\alpha\sigma\gamma\delta}\partial_{\beta}\beta^{\sigma}-X_{\alpha\beta\sigma\delta}\partial_{\gamma}\beta^{\sigma}\nonumber\\ & -X_{\alpha\beta\gamma\sigma}\partial_{\gamma}\beta^{\sigma}-T_{2\sigma\alpha\beta}\partial_{\delta}\partial_{\gamma}\beta^{\sigma}\\
\delta x_{\alpha\beta\gamma}&=-\beta^{\sigma}\partial_{\sigma}X_{\alpha\beta\gamma}-X_{\sigma\beta\gamma}\partial_{\alpha}\beta^{\sigma}-X_{\alpha\sigma\gamma}\partial_{\beta}\beta^{\sigma}-X_{\alpha\beta\sigma}\partial_{\gamma}\beta^{\sigma}-P_{26\alpha\sigma}\partial_{\gamma}\partial_{\beta}\beta^{\sigma}\\
\delta x_{2\alpha\beta}&=-X_{2\gamma\beta}\partial_{\alpha}\beta^{\gamma}-X_{2\gamma\alpha}\partial_{\beta}\beta^{\gamma}-\beta^{\gamma}\partial_{\gamma}X_{2\alpha\beta}-P_{24\gamma}\partial_{\alpha}\partial_{\beta}\beta^{\gamma}\\
\delta t_{2\alpha\beta\gamma}&=-\beta^{\sigma}\partial_{\sigma}T_{2\alpha\beta\gamma}-T_{2\sigma\beta\gamma}\partial_{\alpha}\beta^{\sigma}-T_{2\alpha\sigma\gamma}\partial_{\beta}\beta^{\sigma}-T_{2\alpha\beta\sigma}\partial_{\gamma}\beta^{\sigma}-2P_{22\sigma\alpha}\partial_{\gamma}\partial_{\beta}\beta^{\sigma}\\
\delta y_{5\alpha\beta}&=-A_{5\gamma}\partial_{\alpha}\partial_{\beta}\beta^{\gamma}-Y_{5\alpha\gamma}\partial_{\beta}\beta^{\gamma}-Y_{5\beta\gamma}\partial_{\alpha}\beta^{\gamma}-\beta^{\gamma}\partial_{\gamma}Y_{5\alpha\beta}\\
\delta a_{3\alpha\beta}&=-2X_{2\alpha\beta}-\beta^{\gamma}T_{2\gamma\alpha\beta}+2Y_{5\alpha\beta}\\
\delta \rho_{1\alpha\beta}&= -4X_{\alpha\beta}-X_{\gamma\alpha\beta}\beta^{\gamma}-4X_{2\alpha\beta}\\
\delta t_{\alpha\beta\gamma}&=-4X_{\alpha\sigma\beta\gamma}\beta^{\sigma}-2T_{2\sigma\gamma\beta}\partial_{\alpha}\beta^{\sigma}-2X_{\alpha\beta\gamma}\\
\delta p_{5\alpha\beta}&=-\beta^{\gamma}\partial_{\gamma}P_{5\alpha\beta}-P_{5\gamma\beta}\partial_{\alpha}\beta^{\gamma}-P_{5\gamma\alpha}\partial_{\beta}\beta^{\gamma}\\
\delta\rho_{25\alpha}&=-\beta^{\gamma}\partial_{\gamma}P_{25\alpha}-P_{25\gamma}\partial_{\alpha}\beta^{\gamma} \\
\delta \rho_{26\alpha\beta}&=-\beta^{\gamma}\partial_{\gamma}P_{26\alpha\beta}-P_{26\gamma\beta}\partial_{\alpha}\beta^{\gamma}-P_{26\alpha\gamma}\partial_{\beta}\beta^{\gamma}\\
\delta\chi_{\alpha}&=-\beta^{\gamma}\partial_{\gamma}Q_{\alpha}-Q_{\gamma}\partial_{\alpha}\beta^{\gamma}\\
\delta\chi_{1\alpha}&=2Q_{\alpha}-2\partial_{\alpha}H-2P_{25\alpha}-\beta^{\gamma}P_{26\alpha\gamma}\\
\delta\rho_{10\alpha}&=-4P_{1\alpha}-4P_{25\alpha}-2P_{5\gamma\alpha}\beta^{\gamma}\\
\delta x_{1\alpha\beta}&=-4P_{5\alpha\beta}-2X_{\alpha\gamma\beta}\beta^{\gamma}-2P_{26\beta\gamma}\partial_{\alpha}\beta^{\gamma}\\
\label{eq:rho23-ambig}
\delta\rho_{23}&=-\beta^{\gamma}\partial_{\gamma}P_{23}\\
\delta\rho_{24\alpha}&=-\beta^{\gamma}\partial_{\gamma}P_{24\alpha}-P_{24\gamma}\partial_{\alpha}\beta^{\gamma} \\
\label{eq:c-ambig}
\delta c &=-\beta^{\alpha}\partial_{\alpha}H\\
\delta h_{2}&=-2H-4P_{23}-\beta^{\gamma}P_{24\gamma}\\
\delta \rho_{12}&=-4P_{4}-8P_{23}-P_{25\alpha}\beta^{\alpha}\\
\delta\rho_{13\alpha}&=-2P_{24\gamma}\partial_{\alpha}\beta^{\gamma}-2P_{25\alpha}-2\beta^{\gamma}X_{2\alpha\gamma} \\
\delta \rho_{22\alpha\beta} &= -\beta^{\gamma}\partial_{\gamma}P_{22\alpha\beta}-P_{22\alpha\gamma}\partial_{\beta}\beta^{\gamma}-P_{22\gamma\beta}\partial_{\alpha}\beta^{\gamma}\\
\delta a_{5\alpha}&=-\beta^{\gamma}\partial_{\gamma}A_{5\alpha}-A_{5\gamma}\partial_{\alpha}\beta^{\gamma}\\
\label{eq:a4-ambig}
\delta a_{4\alpha}&=2A_{5\alpha}-2\beta^{\gamma}P_{22\gamma\alpha}-2P_{24\alpha}\\
\delta\rho_{7\alpha}&=-4Y_{\alpha}-4P_{24\alpha}-P_{26\gamma\alpha}\beta^{\gamma}\\
\delta \rho_{21\alpha\beta}&=-2T_{2\alpha\gamma\beta}\beta^{\gamma}-4P_{22\gamma\beta}\partial_{\alpha}\beta^{\gamma}-2P_{26\alpha\beta}\\
\label{eq:a-ambig}
\delta a&=-\beta^{\alpha}\partial_{\alpha}A\\
\delta n&=4A + 2H-A_{5\alpha}\beta^{\alpha}\\
\delta h_1 &= 4H-4Q-\beta^{\alpha}Q_{1\alpha}\\
\delta a_{7\alpha}&=-2A_{5\gamma}\partial_{\alpha}\beta^{\gamma}-2Y_{5\alpha\gamma}\beta^{\gamma}-2Q_{\alpha}+2\partial_{\alpha}H
\end{align}
\end{subequations}

\section{S-theorem: $0+1$D conformal quantum mechanics}
\label{app:1D}
One may wonder whether the formalism that leads to the Weyl anomaly and consistency
conditions can be used for the case of $d=0$. One encounters an immediate obstacle when
attempting this.  There is no immediate generalization of the trace anomaly equation
\eqref{eq:traceAnom}. The problem is that there is no extension of the action integral
that gives invariance under the local version of rescaling transformations, because there
is no extrinsic curvature tensor at our disposal. The naive generalisation of the
Callan-Symanzik equation  specialized to $d=0$,
$H=\beta^{\alpha}\mathcal{O}_{\alpha}$, cannot hold. In fact, for
example, the free particle is a scale invariant system with $H\neq 0$.  

The inverse square potential serves as a test ground for a simple realisation of the
quantum anomaly, where the classical scale symmetry is broken by quantum mechanical
effects~\cite{case} leading to dimensional transmutation {\it i.e,} after renormalization
the quantum system acquires an intrinsic length scale~\cite{gupta, camblong}.  Studies
have been made of non-self-adjointness of the Hamiltonian in the strongly attractive
regime and how to obtain its self-adjoint extension, a procedure that effectively amounts to
renormalisation~\cite{gupta1, griffiths}. The system is also shown to exhibit limit cycle
behaviour in renormalization group flows~\cite{hammer, beane}. This potential appears in
different branches of physics, from nuclear physics~\cite{beane, hammer2} and molecular
physics~\cite{camblong2} to quantum cosmology~\cite{sridip, sridip1, sridip2} and the study of black
holes~\cite{strominger2}.  Given this, it is of interest to understand how quantum effects
break scale symmetry in non-relativist quantum mechanics. We will prove a general theorem
concerning the breaking of scale symmetry.

%
%%%%%%%%%%%%%%%%%
In the quantum mechanical description of a scale invariant system, the Hamiltonian $H$ and
the generator of scale transformations $D$ obey the following commutation relation:
\begin{equation}\label{c}
[D,H]=i z H
\end{equation}
where $z$ is the dynamical exponent of the theory.
We will show an elementary {\it S-Theorem}, that \eqref{c} is incompatible with $H$ being
Hermitian on a domain containing the state\footnote{That is, the action of $D$ on non-zero
  energy eigenstates is well defined.} $D|E\rangle$, where $|E\rangle$ is any non-zero
energy eigenstate. The $S$-Theorem can be used to deduce that classically scale invariant
systems, {\it e.g.,} the inverse square potential, cannot be quantized without loosing
either unitarity or scale invariance if we insist on having bound states with finite
non-zero binding energy.

To prove the theorem, we consider the eigenstates $|E\rangle$ of the Hamiltonian $H$ and take expectation value of the $[D,H]$ in these eigenstates. We have
\begin{equation}
\label{s1}
\langle E| [D,H]| E\rangle = \langle E| DH|E\rangle - \langle E|HD|E\rangle
\end{equation}
Assuming $H$ is hermitian and $D$ is well defined  we have 
\begin{equation} \label{s3}
\langle E| [D,H]| E\rangle = 0
\end{equation}
On the other hand, scale invariance, Eq.~\eqref{c},  implies
\begin{equation}\label{s4}
\langle E| [D,H]| E\rangle = i z \langle E| H| E\rangle \neq 0 
\end{equation}
Comparing \eqref{s3} and \eqref{s4}, proves the  theorem. 
It deserves mentioning that the mismatch is not due to the real part of the quantity $\langle E| \left[D, H\right] |E \rangle$ since,
 \begin{equation}
\text{Re} (\langle E| \left[D, H\right] |E \rangle) =0
\end{equation}
is consistent with 
\begin{equation} \label{cc2}
\text{Re} (\langle E| i zH |E \rangle) =0
\end{equation}
That the mismatch between \eqref{s3} and \eqref{s4} lies in the
imaginary part hints at the fact that $H$ can not be hermitian if we
have scale invariance. We recall that hermiticity of $H$ crucially
depends on vanishing of a boundary term, which is imaginary when we
consider quantities like $\langle E|H|E\rangle$.

For a simple illustration of $S$-theorem consider the free particle
with one degree of freedom, $H=\frac12p^2$ and $D=\frac12(xp+px)-tH$.
Consider first the particle in a finite periodic box with length $L$.
The operator algebra of the free particle holds regardless of the
presence of the periodic boundaries, so the $S$-theorem holds and it
tells us that either $H$ is not hermitian or $D|p\rangle$ is not a
state.  It is instructive to look carefully at the derivation of
\eqref{s3} and \eqref{s4} in this context.  An elementary computation
gives
\begin{equation} 
\label{f1-a}
\langle p | \left( HD|p\rangle\right) - \langle p | \left( DH|p\rangle\right)  = -i p^{2} 
\end{equation}  
which is consistent with the scaling algebra 
\begin{equation}
[D,H]=2i H\,,
\end{equation}
but consistency comes at the expense of rendering $H$ non-hermitian on
a domain which contains the state $D|p\rangle$.  Indeed, for the
periodic box $D|p\rangle$ does not belong in the Hilbert space since
$\langle x | D|p\rangle$ is not periodic. Hence, the apparent loss of
hermiticity is irrelevant as it involves only functions that are not
states.  In the boundary free case ($L\to\infty$) the normalization of
the continuum of energy eigenfunctions is by a Dirac-delta
distribution, and the norm of the functions $\langle x | D|p\rangle$
involves up to two derivatives of the distribution. If we include
these functions in the Hilbert space the Hamiltonian is not hermitian.
On the other hand, if we choose the Hilbert space to be that of square
integrable functions, then $H$ is hermitian but neither
$\langle x | p\rangle$ nor $\langle x | D|p\rangle$ are in the Hilbert
space.

In contrast, consider the inverse square potential problem. For
sufficiently strong attractive potential there are normalizable bound
states $|E\rangle$, and  the state $D|E\rangle$
is properly normalized.  The Hamiltonian is hermitian, but this case
requires reguralisation and renormalization and 
scale symmetry is broken.

This is in fact a statement of a more general result. A corollary of
the $S$-theorem is that we cannot have (properly normalized) bound
states with non-zero energy in a scale invariant system if we insist
on the Hamiltonian being hermitian on the Hilbert space. As in the
previous example, this follows from observing that if there exists a
properly normalized state $|E\rangle$, then $D|E\rangle$ is also a
properly normalized state since the wave-function vanishes
sufficiently fast at infinity. This result is consistent with
representation theory: a discrete spectrum $\{ E_n \} $ cannot form a
representation of a transformation which acts by $ E \to \lambda^z E$
for continuous $\lambda$, (except if the only allowed finite energy
value is $E=0$).  For example, it is well known that for the inverse
square problem in the strongly attractive regime, continuous spectrum
is an illusion since in that regime, $H$ is no more Hermitian. To make
$H$ hermitian, we need to renormalize the problem, breaking the scale
symmetry.

The $S$-theorem can be generalised to to any Hermitian operator $A$ with non zero scaling
dimension $\alpha$, that is,  $[D,A]= i  \alpha A$. The operator
$A$ can not be Hermitian on a domain containing
$D|A\rangle$ where $|A\rangle$ is the eigenstate of operator $A$. In particular, if we
want $A$ to be hermitian on a Hilbert space, $\mathcal{L}^{2}$, then the state
$D|A\rangle$ can not belong to $\mathcal{L}^{2}$. For example, $A$ can be the
momentum operator $p$, which is hermitian on a rigged Hilbert space and has a non-zero
scaling dimension. This generalized $S$-theorem implies that $D|p\rangle$ can not  belong to
the rigged Hilbert space, which is indeed the case.


\begin{thebibliography}{99}
\bibitem{R} C. A. Regal, M. Greiner, and D. S. Jin, \htmladdnormallink{Phys. Rev. Lett. 92, 040403 (2004)}{http://dx.doi.org/10.1103/PhysRevLett.92.040403}.
\bibitem{Z} M. W. Zwierlein et al., \htmladdnormallink{Phys. Rev. Lett. 92, 120403 (2004)}{http://dx.doi.org/10.1103/PhysRevLett.92.120403}.

\bibitem{misc} D. Eagles, Phys.Rev. 186, 456 (1969).
A. Leggett, Diatomic molecules and cooper pairs, in Modern Trends in the Theory of Condensed
Matter, edited by A. Pekalski and J. Przystawa,  Lecture Notes in Physics Vol. 115, pp. 13-27,
Springer Berlin Heidelberg, 1980.\\
A. J. Leggett, J.~Phys. (Paris) Colloq. 41, C7 (1980).\\
P. Nozieres and S. Schmitt-Rink, J.~Low.~Temp.~Phys. 59, 195 (1985).



\bibitem{kaplan}D. B. Kaplan, M. J. Savage, and M. B. Wise, \htmladdnormallink{Phys.~Lett.~B {\bf 424}, 390 (1998)}{http://dx.doi.org/10.1016/S0370-2693(98)00210-X}, \htmladdnormallink{arXiv:nucl-th/9801034}{http://arxiv.org/abs/nucl-th/9801034}.\\
D. B. Kaplan, M. J. Savage, and M. B. Wise, \htmladdnormallink{Nucl.Phys. B534, 329 (1998)}{http://dx.doi.org/10.1016/S0550-3213(98)00440-4}, \htmladdnormallink{arXiv:nucl-th/9802075}{http://arxiv.org/abs/nucl-th/9802075}.

\bibitem{roberts}J. L. Roberts et al., \htmladdnormallink{Phys. Rev. Lett. {\bf 81}, 5109 (1998)}{http://dx.doi.org/10.1103/PhysRevLett.81.5109}.

\bibitem{chin}C. Chin, V. Vuleti\'c, A. J. Kerman, and S. Chu, \htmladdnormallink{Phys. Rev. Lett. {\bf 85}, 2717 (2000)}{http://dx.doi.org/10.1103/PhysRevLett.85.2717}.\\
P. J. Leo, C. J. Williams, and P. S. Julienne, \htmladdnormallink{Phys. Rev. Lett. {\bf 85}, 2721 (2000)}{http://dx.doi.org/10.1103/PhysRevLett.85.2721}.

\bibitem{loftus} T. Loftus, C. A. Regal, C. Ticknor, J. L. Bohn, and D. S. Jin, \htmladdnormallink{Phys. Rev. Lett. {\bf 88}, 173201 (2002)}{http://dx.doi.org/10.1103/PhysRevLett.88.173201}.

%\cite{Lee:1969fy}
\bibitem{Lee:1969fy} 
  T.~D.~Lee and G.~C.~Wick,
  %``Negative Metric and the Unitarity of the S Matrix,''
  \htmladdnormallink{Nucl.\ Phys.\ B {\bf 9}, 209 (1969)}
  {http://dx.doi.org/10.1016/0550-3213(69)90098-4}
  %%CITATION = doi:10.1016/0550-3213(69)90098-4;%%
  %398 citations counted in INSPIRE as of 07 Apr 2016

%\cite{Lee:1970iw}
\bibitem{Lee:1970iw} 
  T.~D.~Lee and G.~C.~Wick,
  %``Finite Theory of Quantum Electrodynamics,''
  \htmladdnormallink{Phys.\ Rev.\ D {\bf 2}, 1033 (1970)}
  {http://dx.doi.org/10.1103/PhysRevD.2.1033}
  %%CITATION = doi:10.1103/PhysRevD.2.1033;%%
  %298 citations counted in INSPIRE as of 07 Apr 2016


%\cite{Grinstein:2007mp}
\bibitem{Grinstein:2007mp} 
  B.~Grinstein, D.~O'Connell and M.~B.~Wise,
  %``The Lee-Wick standard model,''
  \htmladdnormallink{Phys.\ Rev.\ D {\bf 77}, 025012 (2008)}
  {http://dx.doi.org/10.1103/PhysRevD.77.025012},
  \htmladdnormallink{arXiv:0704.1845 [hep-ph]}{http://arxiv.org/abs/0704.1845}.
  %%CITATION = doi:10.1103/PhysRevD.77.025012;%%
  %140 citations counted in INSPIRE as of 07 Apr 2016

%\cite{Grinstein:2008bg}
\bibitem{Grinstein:2008bg} 
  B.~Grinstein, D.~O'Connell and M.~B.~Wise,
  %``Causality as an emergent macroscopic phenomenon: The Lee-Wick O(N) model,''
  \htmladdnormallink{Phys.\ Rev.\ D {\bf 79}, 105019 (2009)}
  {http://dx.doi.org/10.1103/PhysRevD.79.105019},
  \htmladdnormallink{arXiv:0805.2156 [hep-th]}{http://arxiv.org/abs/0805.2156}.
  %%CITATION = doi:10.1103/PhysRevD.79.105019;%%
  %38 citations counted in INSPIRE as of 07 Apr 2016


%\cite{Hawking:2001yt}
\bibitem{Hawking:2001yt} 
  S.~W.~Hawking and T.~Hertog,
  %``Living with ghosts,''
  \htmladdnormallink{Phys.\ Rev.\ D {\bf 65}, 103515 (2002)}
  {http://dx.doi.org/10.1103/PhysRevD.65.103515},
  \htmladdnormallink{hep-th/0107088}{http://arxiv.org/abs/hep-th/0107088}.
  %%CITATION = doi:10.1103/PhysRevD.65.103515;%%
  %148 citations counted in INSPIRE as of 07 Apr 2016

\bibitem{horava1} P. Horava, \htmladdnormallink{Phys. Rev. D {\bf 79} 084008 (2009)}{http://dx.doi.org/10.1103/PhysRevD.79.084008}, \htmladdnormallink{arXiv: 0901.37752 [hep-th]}{http://arxiv.org/abs/0901.3775}
%\bibitem{horova2} P. Horava, \htmladdnormallink{JHEP {\bf 0903}, 020
%    (2009)}{10.1088/1126-6708/2009/03/020},
%  \htmladdnormallink{arXiv:0812.4287
%    [hep-th]}{http://lanl.arxiv.org/abs/0812.4287}


%\cite{Anselmi:2007ri}
\bibitem{Anselmi:2007ri} 
  D.~Anselmi and M.~Halat,
  %``Renormalization of Lorentz violating theories,''
  \htmladdnormallink{Phys.\ Rev.\ D {\bf 76}, 125011 (2007)}
  {http://dx.doi.org/10.1103/PhysRevD.76.125011},
  \htmladdnormallink{arXiv:0707.2480 [hep-th]}{http://arxiv.org/abs/0707.2480}
  %%CITATION = doi:10.1103/PhysRevD.76.125011;%%
  %91 citations counted in INSPIRE as of 07 Apr 2016


%\cite{Anselmi:2008ry}
\bibitem{Anselmi:2008ry} 
  D.~Anselmi,
  %``Weighted scale invariant quantum field theories,''
  \htmladdnormallink{JHEP {\bf 0802}, 051 (2008)}
  {http://dx.doi.org/10.1088/1126-6708/2008/02/05},
  \htmladdnormallink{arXiv:0801.1216 [hep-th]}{http://arxiv.org/abs/0801.1216}.
  %%CITATION = doi:10.1088/1126-6708/2008/02/051;%%
  %44 citations counted in INSPIRE as of 07 Apr 2016

%\cite{Anselmi:2008bq}
\bibitem{Anselmi:2008bq} 
  D.~Anselmi,
  %``Weighted power counting and Lorentz violating gauge theories. I. General properties,''
  \htmladdnormallink{Annals Phys.\  {\bf 324}, 874 (2009)}
  {http://dx.doi.org/10.1016/j.aop.2008.12.005},
  \htmladdnormallink{arXiv:0808.3470 [hep-th]}{http://arxiv.org/abs/0808.3470}.
  %%CITATION = doi:10.1016/j.aop.2008.12.005;%%
  %73 citations counted in INSPIRE as of 07 Apr 2016

%\cite{Anselmi:2008bs}
\bibitem{Anselmi:2008bs} 
  D.~Anselmi,
  %``Weighted power counting and Lorentz violating gauge theories. II. Classification,''
  \htmladdnormallink{Annals Phys.\  {\bf 324}, 1058 (2009)}
 {http://dx.doi.org/10.1016/j.aop.2008.12.007},
  \htmladdnormallink{arXiv:0808.3474 [hep-th]}{http://arxiv.org/abs/0808.3474}.
  %%CITATION = doi:10.1016/j.aop.2008.12.007;%%
  %69 citations counted in INSPIRE as of 07 Apr 2016

%\cite{Baggio:2011ha}
\bibitem{Baggio:2011ha} 
  M.~Baggio, J.~de Boer and K.~Holsheimer,
  %``Anomalous Breaking of Anisotropic Scaling Symmetry in the Quantum Lifshitz Model,''
  \htmladdnormallink{JHEP {\bf 1207}, 099 (2012)}
  {http://dx.doi.org/10.1007/JHEP07(2012)099},
  \htmladdnormallink{arXiv:1112.6416 [hep-th]}{http://arxiv.org/abs/1112.6416}.
  %%CITATION = doi:10.1007/JHEP07(2012)099;%%
  %36 citations counted in INSPIRE as of 23 Mar 2016

\bibitem{arav1} I.~Arav, S.~Chapman and Y.~Oz, \htmladdnormallink{JHEP {\bf 1502} (2015) 078}{http://dx.doi.org/10.1007/JHEP02(2015)078},
\htmladdnormallink{arXiv: 1410.5831 [hep-th]}{https://arxiv.org/abs/1410.5831}.

\bibitem{arav2} I.~Arav, S.~Chapman and Y.~Oz, \htmladdnormallink{arXiv: 1601.06975 [hep-th]}{https://arxiv.org/abs/1601.06795}.

\bibitem{auzzi} R.~Auzzi, S.~Baiguer, G.~Nardelli , \htmladdnormallink{JHEP {\bf 02} (2016) 003}{http://dx.doi.org/10.1007/JHEP02(2016)003},  \htmladdnormallink{JHEP {\bf 02} (2016) 177}{http://dx.doi.org/10.1007/JHEP02(2016)177}, \htmladdnormallink{arXiv: 1511.08150[hep-th]}{https://arxiv.org/abs/1511.08150}.

\bibitem{kristan} K.~Jensen, \htmladdnormallink{arXiv: 1412.7750 [hep-th]}{https://arxiv.org/abs/1412.7750}.

%\cite{Griffin:2011xs}
\bibitem{Griffin:2011xs} 
  T.~Griffin, P.~Horava and C.~M.~Melby-Thompson,
  %``Conformal Lifshitz Gravity from Holography,''
  \htmladdnormallink{JHEP {\bf 1205}, 010 (2012)}
  {http://dx.doi.org/10.1007/JHEP05(2012)010},
  \htmladdnormallink{arXiv:1112.5660 [hep-th]}{http://arxiv.org/abs/1112.5660}.
  %%CITATION = doi:10.1007/JHEP05(2012)010;%%
  %50 citations counted in INSPIRE as of 23 Mar 2016

%\cite{Osborn:1991gm}
\bibitem{Osborn:1991gm} 
  H.~Osborn,
  %``Weyl consistency conditions and a local renormalization group equation for general renormalizable field theories,''
  \htmladdnormallink{Nucl.\ Phys.\ B {\bf 363}, 486 (1991)}
  {http://dx.doi.org/10.1016/0550-3213(91)80030-P}
  %%CITATION = doi:10.1016/0550-3213(91)80030-P;%%
  %136 citations counted in INSPIRE as of 23 Mar 2016
  
\bibitem{Zamo} A. Zamolodchikov, ``Irreversibility of the Flux of the Renormalization Group in a 2D Field
Theory'', \htmladdnormallink{JETP Lett. {\bf 43}, 730 (1986)}{http://www.jetpletters.ac.ru/ps/1413/article_21504.shtml}.

\bibitem{Osborn2}I. Jack \& H. Osborn, ``Analogs for the $c$ theorem for four-dimensional
  renormalizable field theories'', \htmladdnormallink{Nucl.~Phys.~B {\bf 343}, 647 (1990)}{http://dx.doi.org/10.1016/0550-3213(90)90584-Z}.
 


%\cite{Cardy:1988cwa}
\bibitem{Cardy:1988cwa} 
  J.~L.~Cardy,
  %``Is There a c Theorem in Four-Dimensions?,''
  \htmladdnormallink{Phys.\ Lett.\ B {\bf 215}, 749 (1988)}
  {http://dx.doi.org/10.1016/0370-2693(88)90054-8}
  %%CITATION = doi:10.1016/0370-2693(88)90054-8;%%
  %296 citations counted in INSPIRE as of 22 Mar 2016
  
  %\cite{Grinstein:2013cka}
\bibitem{Grinstein:2013cka} 
  B.~Grinstein, A.~Stergiou and D.~Stone,
  %``Consequences of Weyl Consistency Conditions,''
  \htmladdnormallink{JHEP {\bf 1311}, 195 (2013)}
  {http://dx.doi.org/10.1007/JHEP11(2013)195},
  \htmladdnormallink{arXiv:1308.1096 [hep-th]}{http://arxiv.org/abs/1308.1096}.
  %%CITATION = doi:10.1007/JHEP11(2013)195;%%
  %19 citations counted in INSPIRE as of 22 Mar 2016

%\cite{Grinstein:2014xba}
\bibitem{Grinstein:2014xba} 
  B.~Grinstein, D.~Stone, A.~Stergiou and M.~Zhong,
  %``Challenge to the $a$ Theorem in Six Dimensions,''
  \htmladdnormallink{Phys.\ Rev.\ Lett.\  {\bf 113}, no. 23, 231602 (2014)}
  {http://dx.doi.org/10.1103/PhysRevLett.113.231602},
  \htmladdnormallink{arXiv:1406.3626 [hep-th]}{http://arxiv.org/abs/1406.3626}.
  %%CITATION = doi:10.1103/PhysRevLett.113.231602;%%
  %14 citations counted in INSPIRE as of 23 Mar 2016

%\cite{Grinstein:2015ina}
\bibitem{Grinstein:2015ina} 
  B.~Grinstein, A.~Stergiou, D.~Stone and M.~Zhong,
  %``Two-loop renormalization of multiflavor $\phi^3$ theory in six dimensions and the trace anomaly,''
  \htmladdnormallink{Phys.\ Rev.\ D {\bf 92}, no. 4, 045013 (2015)}
  {http://dx.doi.org/10.1103/PhysRevD.92.045013},
  \htmladdnormallink{arXiv:1504.05959 [hep-th]}{http://arxiv.org/abs/1504.05959}.
  %%CITATION = doi:10.1103/PhysRevD.92.045013;%%
  %6 citations counted in INSPIRE as of 23 Mar 2016

%\cite{Osborn:2015rna}
\bibitem{Osborn:2015rna} 
  H.~Osborn and A.~Stergiou,
  %``Structures on the Conformal Manifold in Six Dimensional Theories,''
  \htmladdnormallink{JHEP {\bf 1504}, 157 (2015)}
  {http://dx.doi.org/10.1007/JHEP04(2015)157},
  \htmladdnormallink{arXiv:1501.01308 [hep-th]}{http://arxiv.org/abs/1501.01308}.
  %%CITATION = doi:10.1007/JHEP04(2015)157;%%
  %12 citations counted in INSPIRE as of 23 Mar 2016

%\cite{Stergiou:2016uqq}
\bibitem{Stergiou:2016uqq} 
  A.~Stergiou, D.~Stone and L.~G.~Vitale,
  %``Constraints on Perturbative RG Flows in Six Dimensions,''
  \htmladdnormallink{arXiv:1604.01782 [hep-th]}{http://arxiv.org/abs/1604.01782}.
  %%CITATION = ARXIV:1604.01782;%%
  %1 citations counted in INSPIRE as of 28 Apr 2016


%\cite{Adam:2009gq}
\bibitem{Adam:2009gq} 
  I.~Adam, I.~V.~Melnikov and S.~Theisen,
  %``A Non-Relativistic Weyl Anomaly,''
  \htmladdnormallink{JHEP {\bf 0909}, 130 (2009)}
  {http://dx.doi.org/10.1088/1126-6708/2009/09/130},
  \htmladdnormallink{arXiv:0907.2156 [hep-th]}{http://arxiv.org/abs/0907.2156}.
  %%CITATION = doi:10.1088/1126-6708/2009/09/130;%%
  %24 citations counted in INSPIRE as of 23 Mar 2016

%\cite{Gomes:2011di}
\bibitem{Gomes:2011di} 
  P.~R.~S.~Gomes and M.~Gomes,
  %``On Ward Identities in Lifshitz-like Field Theories,''
  \htmladdnormallink{Phys.\ Rev.\ D {\bf 85}, 065010 (2012)}
  {http://dx.doi.org/10.1103/PhysRevD.85.065010}, 
  \htmladdnormallink{arXiv:1112.3887 [hep-th]}{http://arxiv.org/abs/1112.3887}.
  %%CITATION = doi:10.1103/PhysRevD.85.065010;%%
  %17 citations counted in INSPIRE as of 23 Mar 2016
  
 

%\cite{Fortin:2011ks}
\bibitem{Fortin:2011ks}
  J.~F.~Fortin, B.~Grinstein and A.~Stergiou,
  %``Scale without Conformal Invariance: An Example,''
  \htmladdnormallink{Phys.\ Lett.\ B {\bf 704} (2011) 74}
  {http://dx.doi.org/10.1016/j.physletb.2011.08.060},
  \htmladdnormallink{arXiv:1106.2540 [hep-th]}{http://arxiv.org/abs/1106.2540}.
  %%CITATION = doi:10.1016/j.physletb.2011.08.060;%%
  %39 citations counted in INSPIRE as of 23 Mar 2016


%\cite{Fortin:2011sz}
\bibitem{Fortin:2011sz}
  J.~F.~Fortin, B.~Grinstein and A.~Stergiou,
  %``Scale without Conformal Invariance: Theoretical Foundations,''
  \htmladdnormallink{JHEP {\bf 1207} (2012) 025}
  {http://dx.doi.org/10.1007/JHEP07(2012)025},
  \htmladdnormallink{arXiv:1107.3840 [hep-th]}{http://arxiv.org/abs/1107.3840}.
  %%CITATION = doi:10.1007/JHEP07(2012)025;%%
  %32 citations counted in INSPIRE as of 23 Mar 2016

%\cite{Fortin:2012ic}
\bibitem{Fortin:2012ic}
  J.~F.~Fortin, B.~Grinstein and A.~Stergiou,
  %``Scale without Conformal Invariance at Three Loops,''
  \htmladdnormallink{JHEP {\bf 1208} (2012) 085}
  {http://dx.doi.org/10.1007/JHEP08(2012)085},
  \htmladdnormallink{arXiv:1202.4757 [hep-th]}{http://arxiv.org/abs/1202.4757}.
  %%CITATION = doi:10.1007/JHEP08(2012)085;%%
  %22 citations counted in INSPIRE as of 23 Mar 2016

%\cite{Fortin:2012cq}
\bibitem{Fortin:2012cq}
  J.~F.~Fortin, B.~Grinstein and A.~Stergiou,
  %``Limit Cycles in Four Dimensions,''
  \htmladdnormallink{JHEP {\bf 1212} (2012) 112}
  {http://dx.doi.org/10.1007/JHEP12(2012)112},
  \htmladdnormallink{arXiv:1206.2921 [hep-th]}{http://arxiv.org/abs/1206.2921}.
  %%CITATION = doi:10.1007/JHEP12(2012)112;%%
  %21 citations counted in INSPIRE as of 23 Mar 2016


%\cite{Fortin:2012hc}
\bibitem{Fortin:2012hc}
  J.~F.~Fortin, B.~Grinstein, C.~W.~Murphy and A.~Stergiou,
  %``On Limit Cycles in Supersymmetric Theories,''
  \htmladdnormallink{Phys.\ Lett.\ B {\bf 719} (2013) 170}
  {http://dx.doi.org/10.1016/j.physletb.2012.12.059},
  \htmladdnormallink{arXiv:1210.2718 [hep-th]}{http://arxiv.org/abs/1210.2718}.
  %%CITATION = doi:10.1016/j.physletb.2012.12.059;%%
  %13 citations counted in INSPIRE as of 23 Mar 2016

%\cite{Fortin:2012hn}
\bibitem{Fortin:2012hn} 
  J.~F.~Fortin, B.~Grinstein and A.~Stergiou,
  %``Limit Cycles and Conformal Invariance,''
  \htmladdnormallink{JHEP {\bf 1301}, 184 (2013)}
  {http://dx.doi.org/10.1007/JHEP01(2013)184}
  \htmladdnormallink{arXiv:1208.3674 [hep-th]}{http://arxiv.org/abs/1208.3674}.
  %%CITATION = doi:10.1007/JHEP01(2013)184;%%
  %54 citations counted in INSPIRE as of 22 Mar 2016

\bibitem{Komargodski:2011vj} 
  Z.~Komargodski and A.~Schwimmer,
  %``On Renormalization Group Flows in Four Dimensions,''
  JHEP {\bf 1112}, 099 (2011)
  doi:10.1007/JHEP12(2011)099
  [arXiv:1107.3987 [hep-th]].
  %%CITATION = doi:10.1007/JHEP12(2011)099;%%
  %289 citations counted in INSPIRE as of 20 Oct 2016

%\cite{Luty:2012ww}
\bibitem{Luty:2012ww} 
  M.~A.~Luty, J.~Polchinski and R.~Rattazzi,
  %``The $a$-theorem and the Asymptotics of 4D Quantum Field Theory,''
  JHEP {\bf 1301}, 152 (2013)
  doi:10.1007/JHEP01(2013)152
  [arXiv:1204.5221 [hep-th]].
  %%CITATION = doi:10.1007/JHEP01(2013)152;%%
  %124 citations counted in INSPIRE as of 20 Oct 2016


\bibitem{efimov}
V. Efimov, ``Energy levels arising from resonant two-body forces in a three-body system,''
\htmladdnormallink{Phys. Lett. B {\bf 33}, 563-564 (1970)}{http://dx.doi.org/10.1016/0370-2693(70)90349-7}
%(http://www.sciencedirect.com/science/article/pii/0370269370903497)



\bibitem{PhysRev.47.903}
Thomas, L. H., ``The Interaction Between a Neutron and a Proton and the Structure of ${\mathrm{H}}^{3}$,''
\htmladdnormallink{Phys. Rev.,
{\bf 47}, 903--909 (1935)}{http://dx.doi.org/10.1103/PhysRev.47.903},
  %url = {http://link.aps.org/doi/10.1103/PhysRev.47.903}

\bibitem{PhysRevLett.89.230401}
  {G\l{}azek, Stanis\l{}aw D. and Wilson, Kenneth G.},
``{Limit Cycles in Quantum Theories},''
\htmladdnormallink{Phys. Rev. Lett.,  {89}, {230401} (2002)}
{http://dx.doi.org/10.1103/PhysRevLett.89.230401},
  %url = {http://link.aps.org/doi/10.1103/PhysRevLett.89.230401}


%\cite{Jafferis:2010un}
\bibitem{Jafferis:2010un} 
  D.~L.~Jafferis,
  %``The Exact Superconformal R-Symmetry Extremizes Z,''
  JHEP {\bf 1205}, 159 (2012)
  doi:10.1007/JHEP05(2012)159
  [arXiv:1012.3210 [hep-th]].
  %%CITATION = doi:10.1007/JHEP05(2012)159;%%
  %332 citations counted in INSPIRE as of 20 Oct 2016



  \bibitem{lovelock} D.~Lovelock, ''The Einstein tensor and its generalizations'', \htmladdnormallink{J.~Math.~Phys. {\bf 12}, 498 (1971)}{http://dx.doi.org/10.1063/1.1665613}.
  \bibitem{jackiw00}R.~Jackiw, ``Delta Function Potentials in Two and Three
Dimensions'' in M.A.B. Beg Memorial Volume (World Scientific, Singapore 1991 p. 25)
\bibitem{nishida}Y.~Nishida and D.~T.~Son,
``Non-Relativistic Conformal Field Theories'', \htmladdnormallink{Phys. Rev.
D 76, 086004 (2007)}{http://dx.doi.org/10.1103/PhysRevD.76.086004}, \htmladdnormallink{arXiv: 0706.3746 [hep-th]}{http://arxiv.org/abs/0706.3746}.

\bibitem{jackiw01}R.~Jackiw, 
``Dynamical Symmetry of the Magnetic Monopole'', 
\htmladdnormallink{Ann. Phys. 129, 183 (1980)}{http://dx.doi.org/10.1016/0003-4916(80)90295-X}; 
``Dynamical Symmetry of the Magnetic Vortex'', 
\htmladdnormallink{Ann. Phys. 201, 83 (1990)}{http://dx.doi.org/0.1016/0003-4916(90)90354-Q}.

\bibitem{strominger1}A.~Strominger et. al., ``Lectures on Superconformal Quantum
Mechanics and Multi-Black Hole Moduli Spaces'',
\htmladdnormallink{arXiv: 9911066v4[hep-th]}{http://arxiv.org/abs/hep-th/9911066v4}


\bibitem{jackiw1}C.~Chamon, R.~Jackiw, S. -Y. Pi and L.~Santos, '``Conformal quantum mechanics as the $CFT_1$
dual to $AdS_2$,'' \htmladdnormallink{Phys. Lett. B {\bf 701}, 503 (2011)}{http://dx.doi.org/10.1016/j.physletb.2011.06.023}, \htmladdnormallink{arXiv: 1106.0726v1 [hep-th]}{http://arxiv.org/abs/1106.0726v1}

%\cite{Jackiw:2012ur}
\bibitem{Jackiw:2012ur} 
  R.~Jackiw and S.-Y.~Pi,
  %``Conformal Blocks for the 4-Point Function in Conformal Quantum Mechanics,''
  \htmladdnormallink{Phys.\ Rev.\ D {\bf 86}, 045017 (2012)}{http://dx.doi.org/10.1103/PhysRevD.86.045017},
  \htmladdnormallink{Erratum: [Phys.\ Rev.\ D {\bf 86}, 089905 (2012)]}{http://dx.doi.org/10.1103/PhysRevD.86.089905},
  \htmladdnormallink{arXiv:1205.0443 [hep-th]}{http://arxiv.org/abs/1205.0443}.
  %%CITATION = doi:10.1103/PhysRevD.86.045017, 10.1103/PhysRevD.86.089905;%%
  %12 citations counted in INSPIRE as of 23 Mar 2016

 \bibitem{dAFF} V. de Alfaro, S. Fubini and G. Furlan, ``Conformal Invariance
in Quantum Mechanics'', Nuovo Cim. 34A, 569
(1976)

%\cite{Sen:2011cn}
\bibitem{Sen:2011cn} 
  A.~Sen,
  %``State Operator Correspondence and Entanglement in $AdS_2/CFT_1$,''
  \htmladdnormallink{Entropy {\bf 13}, 1305 (2011)}
  {http://dx.doi.org/10.3390/e13071305},
  \htmladdnormallink{arXiv:1101.4254 [hep-th]}{http://arxiv.org/abs/1101.4254}.
  %%CITATION = doi:10.3390/e13071305;%%
  %46 citations counted in INSPIRE as of 23 Mar 2016

\bibitem{case} K. M. Case, \htmladdnormallink{Phys. Rev. {\bf 80}, 797 (1950)}{http://dx.doi.org/10.1103/PhysRev.80.797}

\bibitem{gupta} K. S. Gupta and S. G. Rajeev, \htmladdnormallink{Phys. Rev. D {\bf 48}, 5940 (1993)}{http://dx.doi.org/10.1103/PhysRevD.48.5940}.

\bibitem{camblong} H .E. Camblong, L. N. Epele, H. Fanchiotti and C. A. G. Canal, \htmladdnormallink{Phys. Rev. Lett. {\bf 85}, 1590 (2000)}{http://dx.doi.org/10.1103/PhysRevLett.85.1590}.

\bibitem{gupta1} B. Basu-Mallick, P. K. Ghosh, K. S. Gupta, Nucl.Phys. B {\bf 659} (2003) 437-457, \htmladdnormallink{arXiv:hep-th/0207040}{https://arxiv.org/abs/hep-th/0207040}.

\bibitem{griffiths} A. M. Essin and D. J. Griffiths, \htmladdnormallink{Am. J. Phys. {\bf 74}, 109  (2005).}{http://dx.doi.org/10.1119/1.2165248}

\bibitem{hammer} H. W. Hammer, B. G. Swingle, \htmladdnormallink{Annals Phys. {\bf 321}, 306 (2006)}{http://dx.doi.org/10.1016/j.aop.2005.04.017}, \htmladdnormallink{arXiv:quant-ph/0503074}{http://arxiv.org/abs/quant-ph/0503074}

\bibitem{beane} S. R. Beane et. al. \htmladdnormallink{Phys. Rev. A {\bf 64}, 042103 (2001)}{http://dx.doi.org/10.1103/PhysRevA.64.042103}

\bibitem{hammer2} P. F. Bedaque, H. W. Hammer, U. van Kolck, \htmladdnormallink{Phys. Rev. Lett. {\bf 82} 463 (1999).} {http://dx.doi.org/10.1103/PhysRevLett.82.463} 

\bibitem{camblong2} H .E. Camblong, L. N. Epele, H. Fanchiotti and C. A. G. Canal, \htmladdnormallink{Phys. Rev. Lett. {\bf 87} 220402 (2001)}{http://dx.doi.org/10.1103/PhysRevLett.87.220402}.

\bibitem{sridip} S. Pal and N. Banerjee, \htmladdnormallink{Phys. Rev. D {\bf 90}, 104001 (2014).}{http://dx.doi.org/10.1103/PhysRevD.90.104001}.

\bibitem{sridip1} S.~Pal and N.~Banerjee, \htmladdnormallink{Phys. Rev. D, {\bf 91}, 044042 (2015)}{http://dx.doi.org/10.1103/PhysRevD.91.044042}, \htmladdnormallink{arXiv: 1411.1167 [gr-qc]}{https://arxiv.org/abs/1411.1167}.

\bibitem{sridip2} S.~Pal, \htmladdnormallink{Class.Quant.Grav. {\bf 33} (2016) no.4, 045007 }{http://dx.doi.org/10.1088/0264-9381/33/4/045007 }, \htmladdnormallink{arXiv: 1504.02912 [gr-qc]}{https://arxiv.org/abs/1504.02912}.

\bibitem{strominger2} A. Strominger, \htmladdnormallink{J. High Energy Phys.{\bf 9802}, 009 (1998)}{http://dx.doi.org/10.1088/1126-6708/1998/02/009}.














\end{thebibliography}
\end{document}